\documentclass[preprint2]{aastex62}
    \usepackage{amsmath,amssymb,graphicx,microtype}

    \received{}
    \revised{}
    \accepted{}
    \submitjournal{ApJ}
    
\begin{document}
        \title{RADYNVERSION: Learning to Invert a Solar Flare Atmosphere with Invertible Neural Networks}
        \correspondingauthor{Chrisopher M.J. Osborne}
        \email{c.osborne.1@research.gla.ac.uk}
        
        \author[0000-0002-2299-2800]{Christopher M.J. Osborne}
        \affiliation{SUPA School of Physics and Astronomy, University of Glasgow, Glasgow, G12 8QQ, Scotland, U.K.}
        
        \author[0000-0003-1589-9365]{John A. Armstrong}
        \affiliation{SUPA School of Physics and Astronomy, University of Glasgow, Glasgow, G12 8QQ, Scotland, U.K.}
        
        \author[0000-0001-9315-7899]{Lyndsay Fletcher}
        \affiliation{SUPA School of Physics and Astronomy, University of Glasgow, Glasgow, G12 8QQ, Scotland, U.K.}
        \affiliation{Rosseland Centre for Solar Physics, University of Oslo, P.O.Box 1029 Blindern, NO-0315 Oslo, Norway}

        \begin{abstract}
        During a solar flare, it is believed that reconnection takes place in the corona followed by fast energy transport to the chromosphere.
        The resulting intense heating strongly disturbs the chromospheric structure, and induces complex radiation hydrodynamic effects.
        Interpreting the physics of the flaring solar atmosphere is one of the most challenging tasks in solar physics.
        Here we present a novel deep learning approach, an invertible neural network, to understanding the chromospheric physics of a flaring solar atmosphere via the inversion of observed solar line profiles in H$\alpha$ and Ca\textsc{ii} $\lambda$8542.
        Our network is trained using flare simulations from the 1D radiation hydrodynamics code RADYN as the expected atmosphere and line profile.
        This model is then applied to single pixels from an observation of an M1.1 solar flare taken with SST/CRISP instrument just after the flare onset.
        The inverted atmospheres obtained from observations provide physical information on the electron number density, temperature and bulk velocity flow of the plasma throughout the solar atmosphere ranging from 0-10~Mm in height.
        The density and temperature profiles appear consistent with the expected atmospheric response, and the bulk plasma velocity provides the gradients needed to produce the broad spectral lines whilst also predicting the expected chromospheric evaporation from flare heating.
        We conclude that we have taught our novel algorithm the physics of a solar flare according to RADYN and that this can be confidently used for the analysis of flare data taken in these two wavelengths.
        This algorithm can also be adapted for a menagerie of inverse problems providing extremely fast ($\sim10\,\mu$s) inversion samples.
        \end{abstract}

        \vspace{3em}
        \section{Introduction} \label{sec:intro}
        The current and next generation of solar observations, with their high spatial, temporal and spectral resolution present a significant analysis challenge, as does the increasing complexity and realism of the models with which the data are confronted. The two go hand-in-hand: ever-increasing resolution reveals observational phenomena that cannot be understood using convenient theoretical simplifications, while the inclusion of `realistic physics' in models (often taken to mean e.g. multi-fluid effects, non-equilibrium processes) motivates observational testing at higher and higher resolution. The challenge of model-data comparison grows accordingly and drives us to seek new approaches. 
        
        This paper deals specifically with combining models and observations to learn about the structure of the solar atmosphere during a solar flare. The underlying motivation for such investigations is to understand how the energy released in a flare is transported through and dissipated in the solar atmosphere, primarily in the solar chromosphere where most of the flare's radiation originates  \citep[appearing mostly in the optical and UV, e.g.][]{Kretzschmar2011,Milligan2014}. However, the route to this is complicated. The observed chromospheric radiation - a combination of optically thin (mostly extreme UV) and optically thick (mostly UV to optical) carries information about the temperature, density and velocity structure of the solar chromosphere, which evolves rapidly with time as it heats. This structure is determined by the pre-flare chromosphere and by the characteristics of the flare energy input. The task is to work out the chromospheric structure from the radiation emitted, and use this to constrain properties of the energy input. The picture is complicated because the heating is very intense - between $10^{10}-10^{12}\rm{erg~cm^{-2}~s^{-1}}$ \citep{Fletcher2007,Krucker2011}, compared to the $\sim 10^7\rm{erg~cm^{-2}~s^{-1}}$ \citep{Withbroe1977} needed to balance radiative losses in the non-flaring chromosphere, and there is abundant evidence for non-thermal particles and flows close to the sound speed, meaning that simplifying assumptions such as hydrostatic or local thermodynamic equilibrium are unlikely to be valid.
        
         We focus here on optically thick emission lines from the upper photosphere and chromosphere. These lines encode information about the atmospheric structure; typically the emergent radiation in the line core is  formed higher up in the atmosphere than in the line wings.
        A number of techniques exist for `inverting' optically-thick line profiles to recover the structure of the atmosphere that emitted them, though most have been developed for the inversion of spectropolarimetric information to include also the magnetic field, which is not our concern at present. These include analytic methods employing the Milne-Eddington approximation for frequency-independent opacity in an LTE atmosphere \citep[e.g.][]{Skumanich1987}, the non-LTE codes NICOLE \citep{SocasNavarro2000} and HAZEL \citep{Ramos2008} and the non-LTE code STiC \citep{DelaCruzRodriguez2018} which can treat multiple atomic species and a complex atmospheric stratification. In essence, these all iterate the output of a forward model towards the observed spectropolarimetric line profiles \citep[note, an alternative approach for solving the inverse problem for the chromospheric temperature structure from an integral form was demonstrated by][]{Metcalf1990}. They have also not been developed with the flare chromosphere in mind, though NICOLE has been used by \cite{Kuridze2017,Kuridze2018} for flares. While non-LTE calculations are included in many codes, hydrostatic equilibrium is uniformly assumed. Instead, the most frequently used approach for flares is forward modeling with codes such as RADYN \citep{Carlsson1992,Carlsson1997,Allred2005,Allred2015} in an attempt to match with observed spectral lines. The energy input to the model is specified according to observed properties when possible (i.e. the energy input by non-thermal electrons deduced from hard X-rays). This approach has produced some notable insights into the properties of the flare chromosphere from both line and continuum emissions \citep[e.g.][]{2015Kuridze,Costa2016,Kowalski2017,Simoes2017}. However, iterating these models towards agreement with observations is not practical, and in some cases reproducing features of the observations pushes the models in ways which are difficult to justify observationally \citep[e.g. the long beam injection times required by][]{Kennedy2015}. Also, while manageable for small samples of data,  this `trial and error' approach cannot realistically be scaled up to take advantage of the high volumes of data from new instruments. Furthermore, in cases where the energy input by non-thermal electrons cannot be constrained because of lack of complementary observations, it is hard to know where to start among the vast range of model possibilities. 
        
        Here we take a different track, exploiting developments in machine learning to efficiently recover RADYN-like atmospheres from spectral line profiles. We design and train an invertible neural network \citep[INN; similar to that introduced in][]{Dinh2017,Ardizzone2018} to learn the output H$\alpha$ and Ca\textsc{ii} 8542~\AA\ spectral lines corresponding to many thousands of RADYN atmospheric solutions, and vice versa. The network proves capable of inverting model RADYN spectral line profiles to generate accurately the corresponding RADYN atmospheric parameters, giving us confidence in its ability to recover reasonable, realistic atmospheres from observed flare spectral data. We demonstrate the method on data taken by the CRISP instrument on the Swedish Solar Telescope \citep{Scharmer2003,Scharmer2008}. The method is fast, producing both atmospheric parameters and a measure of their uncertainties in about 44.7~$\mu$s per measurement on a GPU. This makes application to large datasets feasible.
        
        This initial paper is intended to demonstrate proof of concept, underpinning future in-depth analysis of flares. In Section~\ref{sec:INN} we describe the principles of invertible neural networks, and Section~\ref{sec:training} covers how our network is trained and validated on RADYN models. In Section~\ref{sec:inv} we then present the first inversion using this method of real flare data and end with discussion and conclusions in Section~\ref{sec:disc}.
        
        \section{Invertible Neural Networks (INNs)} \label{sec:INN}
        \begin{figure}[htp]
            \centering
            \includegraphics[width=0.49\textwidth]{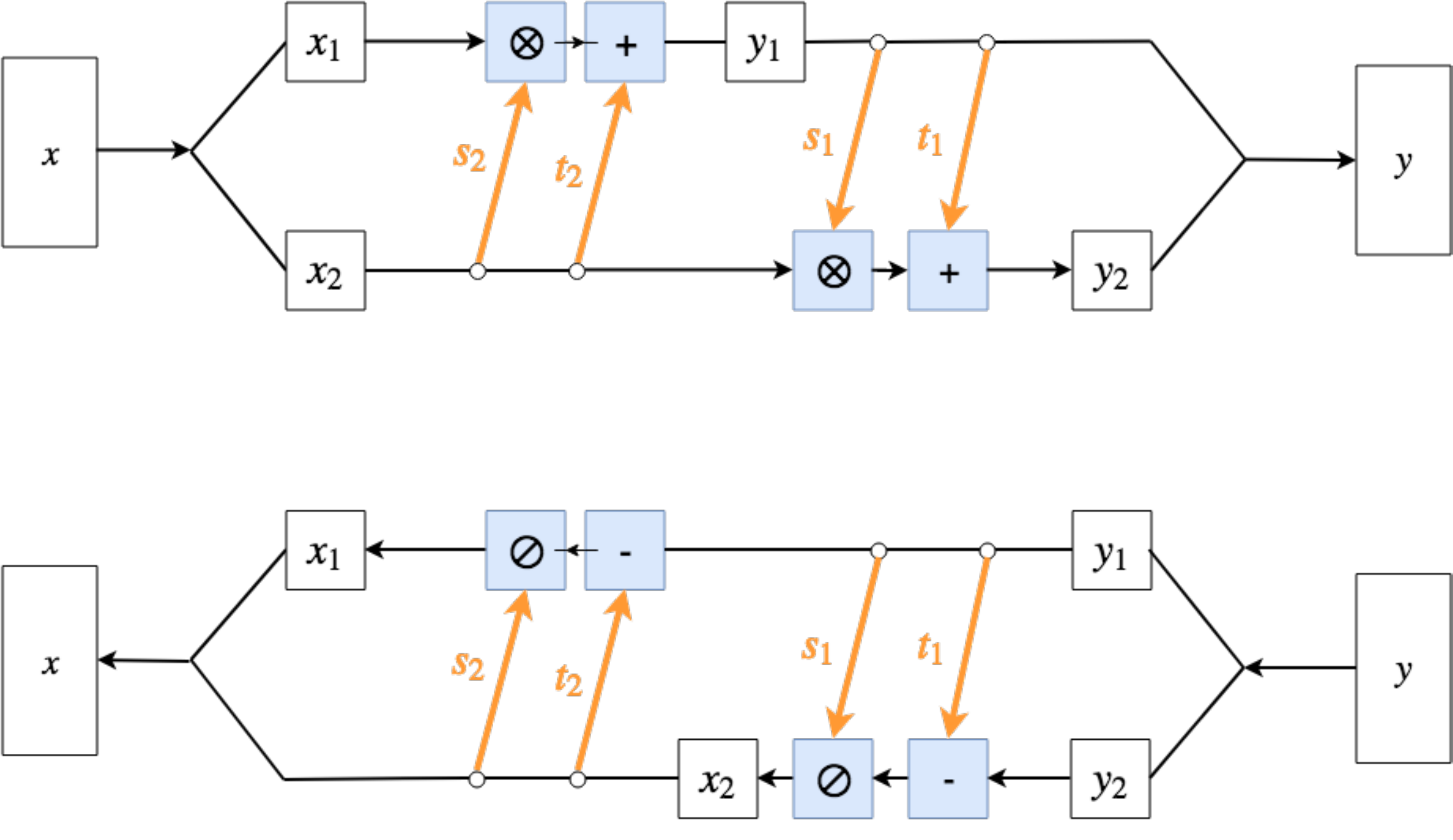}
            \caption{The affine coupling layer showing the affine transformation between input and output for the forward process (top) and the reverse process (bottom). These form the building blocks of our INN as they are easily invertible.}
            \label{fig:affine}
        \end{figure}
        
        An inverse problem is one in which a set of measurements is used to deduce the properties of the system that caused them. It is usually the case that information about the system is missing because of the properties of the medium or the complexity of the physics involved. The example presented in this paper is that of deducing the plasma parameters of the chromosphere which are 3-dimensional quantities, whereas we only observe the chromosphere as two-dimensional images at a given wavelength from an instrument such as the Swedish Solar Telescope CRISP instrument \citep{Scharmer2003,Scharmer2008}.
        We wish to learn about this missing information as it will constrain our model of the physical system producing the observations.
        Formally for any process, there exists a function $y = f (x)$ that maps the input of physical parameters $x$ to the output of observations $y$: this function is known as the {\it forward process}.
       The forward process does not define a bijective function, meaning that we cannot find a unique mapping from the output to the input, i.e. there are many possible $x$ for a single $y$.
        This proves to be important, since a traditional neural network trained on such a problem will only learn to find one of the possible solutions or an average of multiple correct but physically incompatible solutions.
        Furthermore, with a traditional neural network, it is impossible ever to know if the connections being made are the correct ones, as the network is trying to learn an ill-defined problem.
        
         We circumvent this issue in our work by introducing a {\it latent space} $z$ which captures all of the information lost in the forward process \citep[][and references therein]{Dinh2015}. The latent space $z$ represents the space of all information loss in the forward process, such that a sample from the latent space combined with the observation $y$ will be able to be mapped to the correct input parameters $x$. As a result of the introduction of latent variables, we now have a bijective mapping $x \leftrightarrow [y,z]$.
        This means we have transformed the inverse process into a deterministic function (a function which has a definite result for a set of inputs).
        Consequently, sampling different values from the latent space will lead to a sampling of the distribution of the input parameters corresponding to a given output observation.
        This deterministic function $x = g (y,z)$ is thus invertible and we can learn the function $g^{-1}$ as the forward process and $g$ as the inverse process which will track directly where the lost information is obtained from the latent space.
        This is characterised by our network assuming that the latent variables $z$ are drawn from the unit multivariate Gaussian distribution $\mathcal{N} (0, \mathcal{I}_{N})$ for an N-dimensional data space in the reverse direction.
        $g^{-1}$ will populate the true latent space $z_{\text{true}}$ with the information lost in the forward process.
        Our network is then trained in such a way (see Sec. \ref{sec:training}) to learn this mapping from the true latent distribution to the unit Gaussian latent distribution.
        After sufficient training, sampling the unit Gaussian distribution will be equivalent to sampling the true latent distribution since they differ by only a known mapping.
        The choice of drawing from the unit multivariate Gaussian is an arbitrary one.
        It is true that any distribution could be used here but we choose a Gaussian because it is smooth and continuous.
        %\lfq{Is there a reason for assuming that the latent values are drawn from a Gaussian distribution rather than say a uniform distribution? What would be the effect of making a different assumption here? Would the network just learn a different mapping?}
        The architecture we choose to learn this is our invertible neural network.
        
        Invertible neural networks (INNs), like traditional neural networks, are composed of inter-connected layers of neurons which aim to learn a function from input to output.
        The key difference is the composition of the hidden layers between the input and output.
        These take the form of {\it affine coupling} layers \citep{Dinh2015,Dinh2017}.
        Affine coupling layers are simple yet powerful tools.
        By construction, in learning the function from the input to the output with an affine coupling layer we get the inverse function learned for free.
        This is due to the reversibility of the blocks, illustrated in Fig. \ref{fig:affine}.
        We base our layers on the form first presented in \cite{Ardizzone2018}.
        They start by splitting the input $x$ into two equals parts [$x_{1}$,$x_{2}$] and propagating the two halves of the input through the forward direction of the block.
        This leads to $x_{2}$ undergoing an affine transformation before combination with $x_{1}$ to obtain one half of the output $y_{1}$.
        $y_{1}$ is then subject to its own affine transform and combination with $x_{2}$ to get the second half of the output $y_{2}$.
        This is illustrated in the upper panel of Fig. \ref{fig:affine}.
        There is now a simple relation between the input and the output for this layer.
        \begin{align}
            y_{1} &= x_{1} \otimes \exp (s_{2} ( x_{2} )) + t_{2} (x_{2})\\
            y_{2} &= x_{2} \otimes \exp (s_{1} ( y_{1} )) + t_{1} (y_{1})
        \end{align}
        where $\otimes$ denotes the element-wise multiplication of two tensors (which are represented by matrices in our problem) and the functions $s_{i},t_{i}$ are arbitrarily complex and differentiable ($i \in \{1,2\}$).
        After obtaining the pair of outputs [$y_{1}$,$y_{2}$], they are then concatenated to give the total output $y$.
        The inverse of this operation is then simple and we can also map from the output $y$ to the input $x$.
        \begin{align}
            x_{2} &= (y_{2} - t_{1} (y_{1})) \oslash \exp (s_{1} (y_{1}))\\
            x_{1} &= (y_{1} - t_{2} (x_{2})) \oslash \exp (s_{2} (x_{2}))
        \end{align}
        where $\oslash$ denotes the element-wise division of two matrices.
        We have now defined a setup in which the inverse is easily calculable.
        This is extremely useful for inverse problems as it is rarely easy to find the inverse function for a forward model.
        This means that the only problem we now need to deal with is learning what the latent space is to make sure that our network produces the correct inversion, see Sect. \ref{sec:training} for more information.
        Since the functions $s_{i},t_{i}$ do not need to be inverted themselves to calculate the inversion, they can be as complex and arbitrary a function as needed.
        To fill this role we look to fully-connected artificial neural networks (ANNs).
        
        ANNs are widely-known as universal function approximators as they can learn complex classification and regression problems via a method known as backpropagation \citep{Rumelhart1986,Cybenko1989}.
        ANNs are an example of supervised machine learning, meaning that the network is trained on a dataset where the answers to the functions we want to learn are known.
        In backpropagation, the input data is fed through a neural network where linearities and non-linearities are applied to it until it reaches the output where it is compared with the known answers.
        This comparison is then surmised by a loss function which is minimised by changing the values of the weights in each layer of the network to produce a different result \citep{Schmidhuber2015}.
        There have been innumerable successes of ANNs learning complex functions via this method and so we use randomly initialised ANNs as our complex $s_{i}$ and $t_{i}$ functions in the INN.
        
\begin{figure*}[htp]
            \centering
            \includegraphics[width=\textwidth]{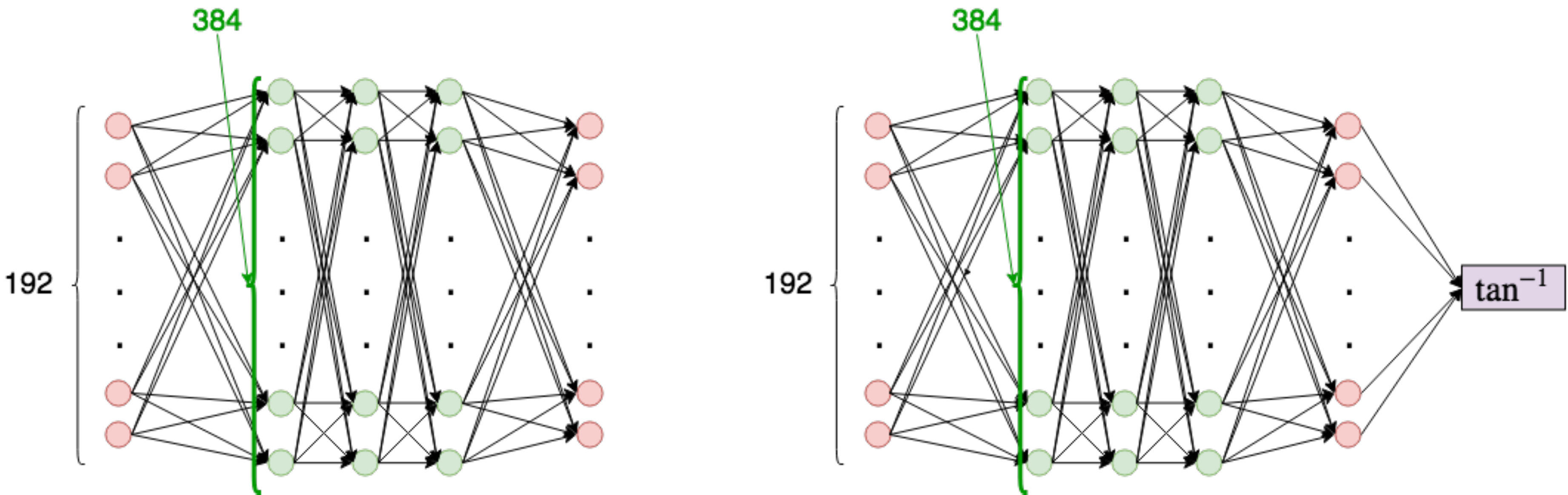}
            \caption{The fully-connected networks for the $t_{i}$ functions (left) and $s_{i}$ functions (right). These are deep neural networks with 4 hidden layers. The network architecture for the $s_{i}$ functions contains a smooth clamping function after output in the form of the inverse tangent. This clamps the output such that the exponential term in our affine transform does not overshadow the linear term (as this would make the linear term null). The input dimension is half the input dimension of the affine coupling layer due to the splitting of the input as shown in Fig. \ref{fig:affine}. The hidden layer depth is then double this.}
            \label{fig:fcn}
        \end{figure*}

        In our network, the functions $s_{i}$ and $t_{i}$ are defined by four layer fully-connected networks (FCN).
        An FCN is a type of ANN where all neurons in the previous layer are connected to all neurons in the current layer.
        The basic architecture for the FCNs utilised in our network is shown in Fig. \ref{fig:fcn}.
        The activation function (the function that determines to what extent the nodes pass on information to the next layer) after the first 3 layers in our deep networks are given by Leaky ReLU (rectified linear unit):
        \begin{equation} \label{eq:lrelu}
            \phi (x) = \text{max} (x, 0.01 x)
        \end{equation}
        with the activation after the fourth given by a ReLU:
        \begin{equation} \label{eq:relu}
            \phi (x) = \text{max} (0, x)
        \end{equation}
        where $x$ is the input (in both cases).
        These activations are used as they are sparse and thus speed up computation.
        Furthermore, ReLU activation and its variants are popular as they are better at avoiding the vanishing gradient problem (when the gradients of the loss are small enough they do not affect the update of the weights leading to the optimiser getting stuck in the loss space).
        The functional forms of $s_{i}$ and $t_{i}$ differ by a clamping inverse tangent function applied at the end of the $s_{i}$ networks.
        This clamping function stops the exponential terms dominating the affine transform whilst still being smooth (i.e. gradients are still easy to calculate).
        These networks are trained as normal via backpropagation (see Sect. \ref{sec:training}) and they learn the optimal representation of the affine transform that will approximate the forward physical model.
        Then this representation is also optimal for the inverse problem as the FCNs apply to the inverse problem too.
        
\begin{figure*}[htp]
            \centering
            \includegraphics[width=\textwidth]{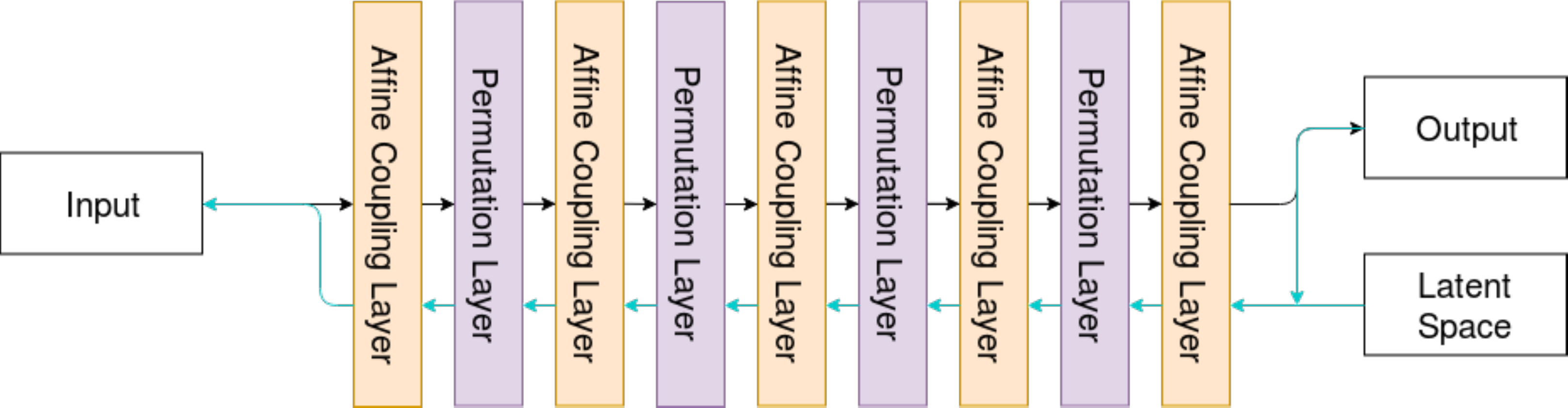}
            \caption{The architecture of our INN. We have five affine coupling layers with a permutation layer sandwiched between two affine coupling layers (four in total). The forward process mapping the input to the output is illustrated by the black arrows. The inverse process mapping a combination of the output and the latent space to the input is illustrated by the cyan arrows.}
            \label{fig:inn}
        \end{figure*}

        Our network is comprised of five stacked affine coupling layers.
        Stacking these layers will allow us to approximate more complex tasks (this is the standard pillar of deep learning \citep{raschka_2015}).
        This means that the network is dependent on 20 deep neural networks to approximate our inverse problem.
        Between each subsequent affine coupling layer, we have what is known as a permutation layer.
        This introduces channel-mixing into our network by permuting the order of the inputs to each new layer.
        Channel-mixing is when the inputs are shuffled into a different order.
       This is done as the input to the affine coupling layers are split in two meaning that if there is no permutation then these two halves remain independent throughout the network.
        %\lf{\bf what is channel mixing and why do we do it?}
        The permutations are done by shuffling the input dimensions of our network in a random but fixed way \citep{Dinh2015,Dinh2017}.
        Each permutation is different from the previous.
        This will increase the generalisation properties of our network.
        The architecture of the INN is shown in Fig. \ref{fig:inn}.
        The flow of the forward model is shown by the black arrows and the flow of the inverse is shown by the cyan arrows.

         \section{Training an INN using synthetic flare data} \label{sec:training}
        
        This Section describes the methods used to train and validate an INN to learn a bijective mapping between atmospheric profiles and two spectral lines.
        The training data consists of synthetic flaring solar atmospheres and spectral line profiles generated from the one-dimensional radiation hydrodynamic model RADYN.
        
        \subsection{Training Data}
        
        The state-of-the-art forward models for simulating the atmospheric response and radiation originating from solar flares are one-dimensional radiation-hydrodynamic models that solve the equations of hydrodynamics coupled with the equations of radiative transfer (outside local thermodynamic equilibrium and statistical equilibrium).
        Amongst these models are RADYN \citep{Carlsson1992,Carlsson1997,Allred2005,Allred2015}, FLARIX \citep{Varady2010,Heinzel2015}, and HYDRAD \citep{Bradshaw2013}.
        Due to the pre-existing grid of RADYN simulations\footnote{Produced by the F-CHROMA project and available from: \url{https://star.pst.qub.ac.uk/wiki/doku.php/public/solarmodels/start}} 
        %\citep{2014Cauzzi} LF I removed this because it pre-dates the RADYN grid
        and its widespread acceptance we have chosen to use RADYN as the forward model for training here.
        These RADYN simulations all start from a modified VAL3C quiet sun atmosphere \citep{Vernazza1981}.
        
        For the simulations in the RADYN grid, the dynamic atmospheric response to an electron beam from a flare is computed, where:
        \begin{itemize}
            \item The distribution of electron energies in this beam is modeled as a power law with variable total energy flux (in the range $3\times10^{10}-1\times10^{12}$ erg cm$^{-2}$).
            \item The beam low energy cut off is \\${E_c=\{10,15,20,25\}}$~keV.
            \item The beam spectral index $\delta=\{3,4,5,6,7,8\}$.
            \item The beam flux is a symmetric triangular pulse, lasting for 20s and peaking at 10s.
            \item The simulation lasts for 50 s with data available every 0.1 s.
        \end{itemize}
        
        Some of the simulations with high total energy, lower values for $E_c$, and higher values for $\delta$ did not complete and therefore are not available in the grid.
        This leaves 81 simulations, with 40,500 total timesteps to use as our training data.
        20\% of these timesteps are separated and used to independently verify the training.
        
        RADYN uses an adaptive spatial grid \citep{Dorfi1987} to accurately represent the atmosphere, but due to the way in which our INN learns shapes this data must be first interpolated onto a fixed, static, grid.
        As we are primarily interested in the chromosphere and transition region, where the plasma parameters vary rapidly in space, we interpolate onto 45 linearly spaced points below 3.5~Mm, with a grid spacing of 79.2~km. Five further points are used to represent the rest of the corona, and these are spaced exponentially from 3.5~Mm up to 10~Mm.
        
        The plasma parameters extracted from the RADYN simulations are the electron density $n_e$ [cm$^{-3}$], the temperature $T$ [K], and velocity $v$ [cm s$^{-1}$] as a function of altitude and time on the interpolated grid. 
        The line profiles from these simulations, for H$\alpha$ 6563~\AA~and Ca 8542~\AA, are each interpolated onto 30 linearly spaced points in wavelength, across wavelength ranges with half-widths 1.4~\AA~and 1.0~\AA~respectively.  The assumption of energy input specifically by an electron beam originating in the corona results in a characteristic Coulomb-collisional energy deposition profile in the chromosphere - determining $n_e, T$ and $v$. For the spectral lines we will use, RADYN calculates both the thermal and the non-thermal (i.e. direct beam-electron electron impact) collisional rates.
       
        To reduce the dynamic range of these profiles and improve the performance of the INN we first map $n_e\mapsto\log_{10}{n_e}$, $T\mapsto\log_{10}{T}$, and $v\mapsto\textrm{sign}(v)\log_{10}(|v| / 10^5 + 1)$.
        For each timestep in each simulation the line profiles are divided by the maximal intensity in each profile, so that the profiles' relative intensities are preserved on a [0--1] scale.
        
        \subsection{Maximum Mean Discrepancy}
        
        Training the INN is made possible by the use of the Maximum Mean Discrepancy (MMD).
        The MMD is a statistic used for computing the distance between two probability distributions based on a set of randomly drawn samples from each distribution by means of a high- or infinite-dimensional space through a non-linear feature mapping.
        Our implementation is explained in depth in Appendix~\ref{App:Mmd} drawing on \citet{Gretton2012} and lectures given at the Machine Learning Summer School 2018\footnote{available at \url{http://www.gatsby.ucl.ac.uk/~gretton/teaching.html}}.
        
        \subsection{Training}
        Our INN is trained similarly to \cite{Ardizzone2018}, and is based on their Framework for Easily Invertible Architectures (FrEIA)\footnote{\url{https://github.com/VLL-HD/FrEIA}}.
        Herein, we provide a more in depth description of the training method and the slight differences in the MMD loss used.
   
        The INN is trained with the preprocessed simulation data alternating forwards and backwards iterations.
        We define the input $x$ as the concatenation of the electron density, temperature and velocity profiles at a certain timestep.
        The output $y$ is the concatenation of the normalised line profiles at this timestep.
        The latent space $z$ is currently defined to be the same length as $x$, although this is still an area of investigation tied to the intrinsic dimensionality of the the problem.
        The output of the INN is then the vector $[z,\,y]$.
        Both the input and output vectors are zero-padded to provide the network blocks with additional dimensionality over which to represent the learnt mapping, and also to fix the input and output to the vectors to the same length, as the structure of the affine coupling layers requires this.
        We will write these zero padded vectors $x_p = [x,\,0,\,0,\,\ldots]$ and $y_p=[z,\,0,\,0,\,\ldots,\,y]$ and in our network these are padded to have a length of 384. 
        
        The forwards and backwards training directions are both constrained by two loss functions.
        A loss function is a function that the neural network optimiser attempts to minimise during training so as to minimise the distance between the output from the ANN and the expected output.
        In the forward direction we apply an L2 loss ($||y-y_\textrm{true}||_2^2$)
        to the zero-padding and line profiles in the generated $y_p$ vector against the expected $y_p$ training vector from the forward model.
        An MMD loss is also applied between batches of $[y,\,z]$ and $[y_\textrm{true},\,\mathcal{N}(0, \mathcal{I}_z)]$. 
        During backpropagation (modification of the weights in the ANN layers guided by the gradients at these nodes) the gradients on the generated $y$ due to the MMD loss are ignored so as to train the neurons learning the mapping from the true latent distribution to the normal distribution without hindering the training of the forward model $x\mapsto y$.
        The convergence of this MMD loss ensures the independence of $z$ from $y$.
        
        The backwards direction is trained similarly.
        The vector of $y_\textrm{true}$ and the latents $z$ generated by the forward iteration is propagated through the network in reverse and an L2 loss is applied between $x_p$ and a zero-padded vector containing $x_\textrm{true}$.
        Another vector of $y_\textrm{true}$ with latents $z$ drawn from the normal distribution are also propagated in reverse and an MMD loss is computed between $x$ and $x_\textrm{true}$.
        This second MMD loss serves to ensure that the distributions of $x$ across the batch look alike (whilst taking into account internal variability within the true distribution).
        
        The kernel used in our MMD loss is the same as that of \cite{Ardizzone2018} and \cite{Tolstikhin2017}, the inverse multiquadric (IMQ) kernel 
        \begin{equation}
            k_\alpha(x,y) = \frac{\alpha^2}{\alpha^2 + ||x-y||_2^2}
        \end{equation}
        
        as it has been found most effective for comparing sample quality in these problems.
        In the example provided by \cite{Ardizzone2018} the kernel used is a sum of IMQ kernels with different $\alpha$ (due to the properties of the Reproducing Kernel Hilbert Space over which the MMD is defined this sum is also a kernel), however we had difficulty isolating a set of values for $\alpha$ that were effective in training the latent distribution to match the expected distribution without dependence on $y$.
        By plotting the MMD for the same $x$ and $y$ samples but different values of $\alpha$ it was found that the biased sample estimate of the MMD between $x$ and $y$ drawn from similar, but perturbed, distributions produced a peak for certain values of $\alpha$.
        We therefore compute the value of $\alpha$ at the turning point of the MMD$^2(\alpha)$ (for which the MMD is maximal) during the training of the net and update $\alpha$ every five epochs.
        This approach is supported by \cite{Sriperumbudur2009}, as the kernel of a family that yields the greatest distinction between the two differing distributions is the one for which the MMD estimate is maximal.
        
        Our INN is trained using the Adam optimiser \citep{Kingma2014} with $\beta_1=\beta_2=0.8$ and $\epsilon=1\times10^{-6}$, where the $\beta$ hyperparameters control the momentum of the first and second moments of the gradients and $\epsilon$ prevents division by zero.
        A hyperparameter is a parameter that is set prior to training, possibly evolving in a predictable fashion, and is not optimised by the training process.
        The values of these parameters are typically determined empirically, and may well not be optimal, but have been chosen to lead to convergence of the model.
        The learning rate $\eta$ (the size of the steps taken in descending the gradient) is initially set to $1.5\times10^{-3}$ and decays by a factor of $\gamma=0.004^{1/1333}$ every 12 epochs, thus for the model presented in this paper, trained for 11400 epochs, the final learning rate is $\eta\approx3.38\times10^{-5}$.
        This model does not appear to be very sensitive to variations in the learning rate and multiple variations of $\gamma$ have been used with success.
        We used a minibatch size of 500, with 20 minibatches per epoch, and the backpropagation took place every minibatch.
        In contrast to traditional training where the model is trained on the entire training set every epoch, and accumulates gradients over the entire training set before backpropagation, minibatch training shows the model multiple small subsets of the data each epoch with gradient accumulation and backpropagation between each of these minibatches.
        
        The two losses computed for each of the forwards and backwards iterations need to be combined into a single loss in each direction for the backpropagation.
        We use this as an opportunity to add additional hyperparameters with which to weight the various losses when combining them. 
        We therefore define three weights $w_\textrm{pred}$, $w_\textrm{latent}$, and $w_\textrm{rev}$.
        Then the loss from the forward process is produced by 
        \begin{equation}
        \textrm{loss}_\textrm{f}=w_\textrm{pred} \textrm{L2}_\textrm{f} + w_\textrm{latent}\textrm{MMD}_\textrm{f},
        \end{equation}
        and the backwards loss by 
        \begin{equation}
        \textrm{loss}_\textrm{b} = 0.5w_\textrm{pred}\textrm{L2}_\textrm{b} + \xi(n)w_\textrm{rev}\textrm{MMD}_\textrm{b},
        \end{equation}
        
        where $f$ and $b$ represent the previously discussed forwards and backwards losses that are combined, $\xi(n)=\left(\textrm{min}\left(\frac{n}{0.4N_\textrm{fade}}, 1\right)\right)^3$ with $n$ the current epoch and $N_\textrm{fade}$ is the number of epochs in the initial training stage. The function $\xi(n)$ helps to avoid the initially large gradients in $\textrm{MMD}_\textrm{b}$ from steering the net away from the correct solution. In practice it was found that this function was not strictly necessary, but improved convergence.
        Additionally, the zero padding was set to $5\times10^{-2}(1-\xi(n))\mathcal{N}(0, 1)$ to increase the activations of these neurons during early training and therefore push their outputs towards zero.
        The exact values of these parameters were determined empirically, but with an emphasis on minimising the L2 losses.
        
        The initial 800 epochs were treated as an initial fade-in stage as $\xi(n)$ grew to 1 and the padding became 0.
        For this phase the loss weightings were set to $w_\textrm{pred}=4000$, $w_\textrm{latent}=900$, and $w_\textrm{rev}=1000$.
        After this initial phase the net was trained in batches of 400 epochs up to 4800 epochs, increasing $w_\textrm{pred}$ by 1000 each batch.
        This process was then repeated with batches of 600 epochs up to a total of 12000 epochs.
        Finally, the model that performed best on the unseen validation set was chosen as the final model.
        This model was trained for 11400 epochs.
        
        \subsection{Validation}\label{sec:validation}
        
        \begin{figure*}[htp]
        \centering
        \includegraphics[height=0.36\textheight]{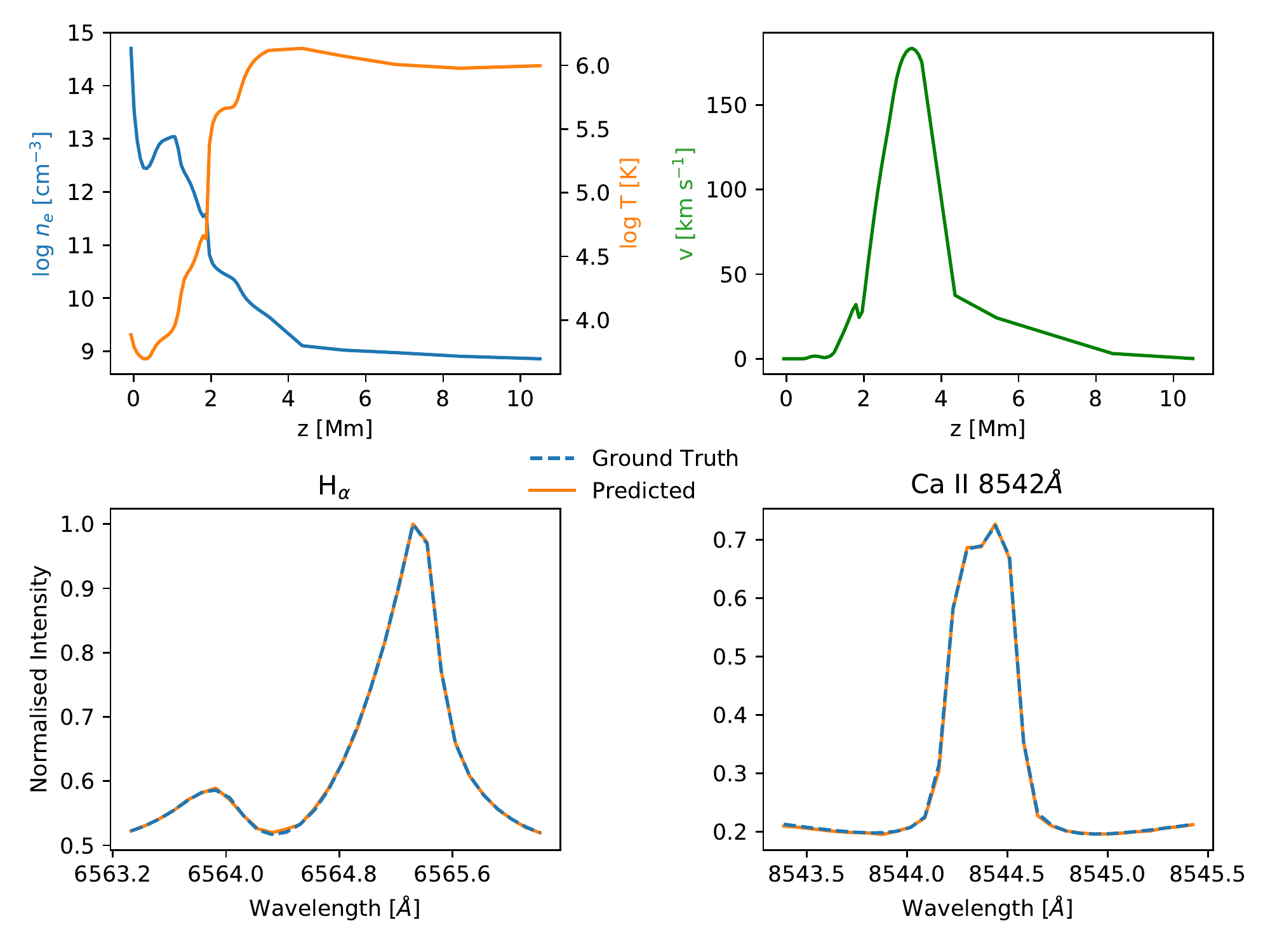}
        \caption{\textbf{Output of the model's forward process on unseen testing data}. The top row shows the atmospheric parameters used as input to the network, and the bottom row shows the output of the model's approximation of the forward process with the true results overlaid with the dashed line. Positive velocities represent upflows.}
        \label{Fig:ForwardVerification}
        \end{figure*}
        
        \begin{figure*}[htp]
        \centering
        \includegraphics[height=0.36\textheight]{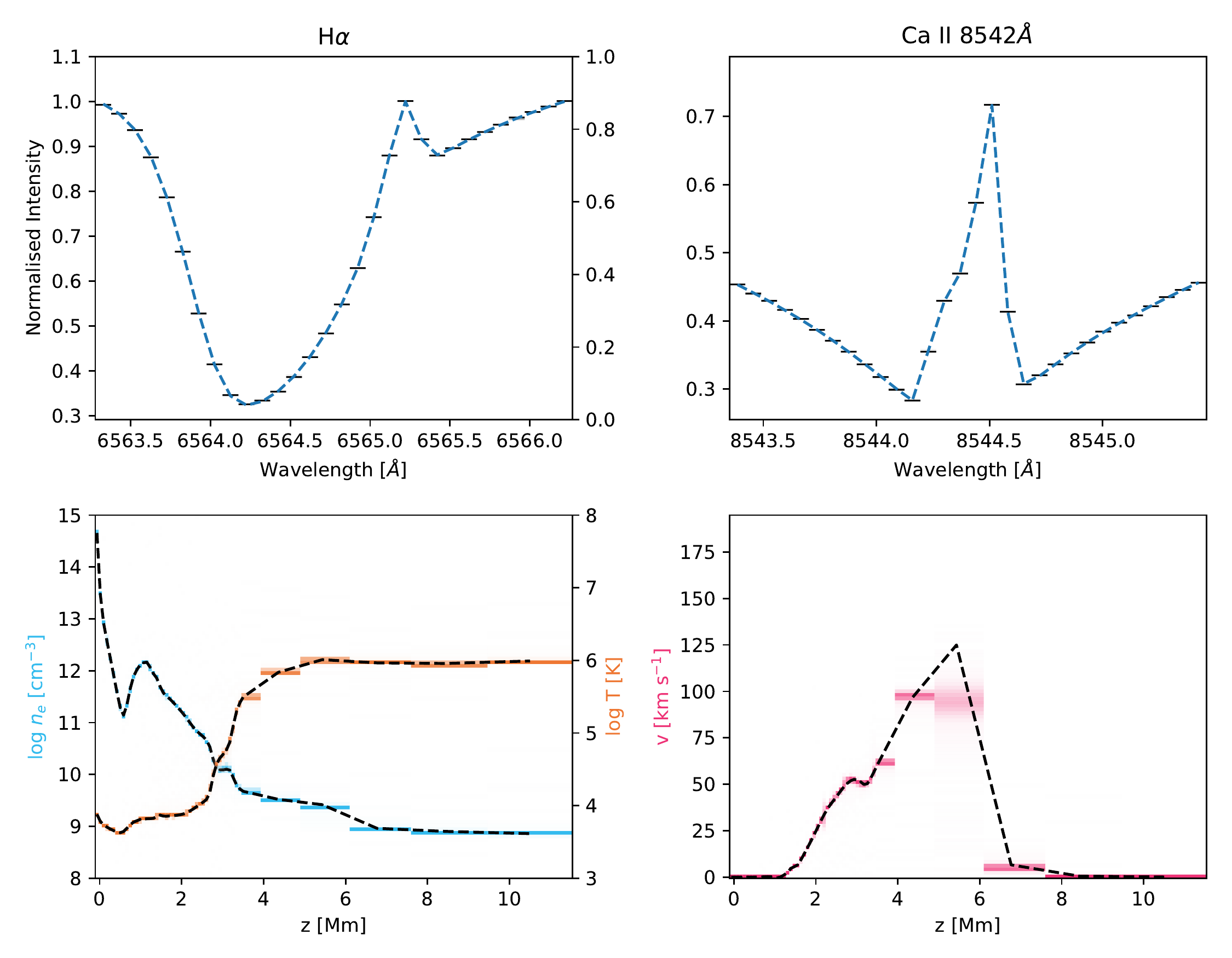}
        \caption{\textbf{Output of the model's inverse process on unseen testing data}. The dashed lines in the top row show the input to the inverse process, that are augmented with a randomly drawn latent space. The two-dimensional histograms in the lower row show the results of each inversion. The dashed lines on the lower row show the expected solution for the inversion. The two-dimensional histograms  (narrow grey bars) in the top row are the result of propagating each atmospheric solution from the inversion through the forward process.}
        \label{Fig:BackwardVerification}
        \end{figure*}
        
        The first stage in validating the training of the model is to test the forward model against ground truths on the unseen testing data. Fig.~\ref{Fig:ForwardVerification} shows the results of the forward model. The top panels are the electron number density, temperature and flow speed from an unseen RADYN snapshot, and the bottom panels compare the `ground truth' RADYN output line profiles with the network's forward process. The mean squared error is $5.73\times10^{-5}$ in the scaled intensity at each wavelength point.
        Note that for all figures in this paper wavelength axes show the wavelength in a vacuum, and positive velocities represent upflows.
        
        It is somewhat more difficult to evaluate the model's ability to reproduce an atmosphere when given the line profiles, due to the aforementioned ambiguity of the problem, as one set of line profiles may have been produced by a variety of atmospheres. 
        To understand the range of solutions, we draw random samples from the latent space multiple times, and use these samples with the line profiles to generate a histogram of atmospheric properties predicted by the INN.
        %%\lfq{I'm having trouble visualising what this actually involves, and what the sampling strategy implies about the latent space properties. The latent space is a vector of values of the same length as the concatenation of the physical input variables, but is it possible to identify  `segments' of the latent space with $n_e$ or $T$ and sample accordingly? Or do you sample across the whole latent space and somehow generate joint distributions for the corresponding variables?}
        Fig~\ref{Fig:BackwardVerification} shows the results and verification of the inversion of data from the unseen testing set.
        On the first row the input line profiles are plotted in dashed blue on top of horizontal bars representing the line profiles calculated using the recovered atmospheric solutions. 
        The recovered solutions are shown in the second row, plotted as two-dimensional coloured histograms representing the probability density of the solution at each altitude node. 
        The regions of highest density in these parameters are therefore the most likely values.
        Superposed on this are the ground truth values for each parameter, plotted as dashed lines.
        The data in the histograms are accumulated for every solution for the atmospheric profile produced from different draws of the latent space and represent 10,000 sampled solutions. 
        
        To better show the range of outlying solutions, all of the histograms were gamma corrected (with $\gamma=0.3$) to reduce contrast. 
        As can be seen from the dashed black line in the lower panels of  Fig.~\ref{Fig:BackwardVerification}, the peak density of the solutions is close to the ground truth, and the narrowness of the histograms show that the solution is well constrained through the atmosphere up to around 3~Mm above the photosphere. However, the spectral lines used do not constrain the problem well in the upper atmosphere, and although the solutions align very well with the ground-truth, the histograms are broader, particularly for the profile of velocity at 4~Mm and above.
        The histograms underneath the input line profiles in the top row of Fig.~\ref{Fig:BackwardVerification} - so narrow as to look like single bars -  are obtained by applying the forward model to each atmosphere produced by the inverse process, and gamma corrected in the same way. They reproduce the input line profiles very closely, demonstrating the self-consistency of the model's solutions.
        
        \section{Single-pixel inversion of real flare data} \label{sec:inv}
        %\lfq{In addition to the bits in red, I did some re-organisation here as the information seemed to me to appear in the wrong order, in places.}
        \begin{figure*}[htp]
            \includegraphics[width=\textwidth]{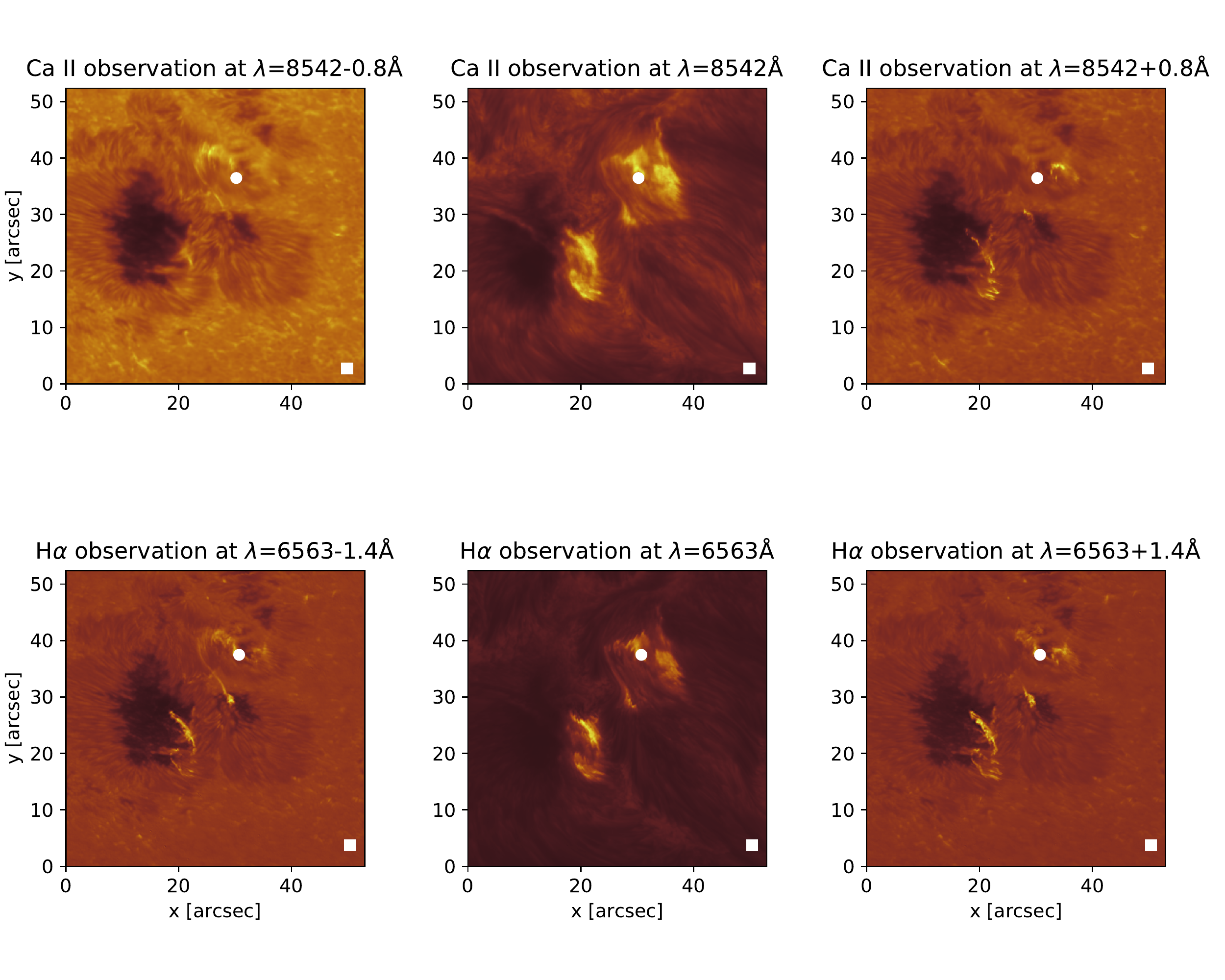}
            \caption{The observations of the M1.1 two-ribbon solar flare from AR12157 on 2014/09/06. These images are from just after the onset of the flare at 16:56:13UTC. The top row shows images taken in the Ca\textsc{ii} 8542\AA{} band with the left panel showing the blue line wing, the middle panel showing the line core and the right panel showing the red line wing. The bottom row shows images taken in the H$\alpha$ band following the same convention as for Ca\textsc{ii}. We select two pixels for our inversion test: one on the flare ribbon (circle) and one off the flare ribbon (square). These points are plotted on top of the images in each panel.}
            \label{fig:images}
        \end{figure*}
        
       We have demonstrated above that the INN has successfully learned the synthetic flare model from RADYN.
       The next step is to apply our learned model to real spectroscopic data, with the intention of characterising the atmosphere that produced it, and eventually learning about the physics of a flaring chromosphere.
        As our problem is only defined in H${\alpha}$ and Ca\textsc{ii} 8542~\AA{} and these are mostly formed in the chromosphere (cores) and the upper photosphere (wings), we will focus specifically on our results for atmospheric parameters below around $z \approx 2$~Mm.
        We do not attach much significance to the results from the small number of points in the corona.
        %\lfq{would it then make sense in the Figures to zoom in on the region where you think the inversions are more reliable, i.e. up to 4Mm or so?}
        
        The flare data we use is from the M1.1 two-ribbon solar flare SOL20140906T17:09 which occurred in NOAA AR12157 with heliocentric coordinates (-732$\arcsec$,\,-302$\arcsec$). Data was taken by the CRisp Imaging SpectroPolarimeter \citep[CRISP;][]{2006Scharmer,Scharmer2008} mounted on the Swedish 1-m Solar Telescope \citep[SST;][]{Scharmer2003} on La Palma.
        CRISP produced imaging spectroscopy data in both H$\alpha$ and Ca\textsc{ii}.
        The H$\alpha$ data consists of 15 wavelength positions sampled at intervals of 200~m\AA{} from the line core, and the Ca\textsc{ii} data consists of 25 wavelength positions sampled at intervals of 100~m\AA{} from the line core.
        The cadence of these observations is 11.54~s with spatial sampling of 0.057$\arcsec\,\text{px}^{-1}$ (giving a spatial resolution of 0.114$\arcsec$).
        %This implies that most of our observations are diffraction-limited. \lf{Only if MOMFBD is doing better than 0.1", and how would we tell?}
        The dataset is open access and available from the F-CHROMA solar flare database \citep{2014Cauzzi}\footnote{\url{https://star.pst.qub.ac.uk/wiki/doku.php/public/solarflares/start}} where it has been pre-processed and reconstructed using Multi-Object Multi-Frame Blind Deconvolution (MOMFBD; \cite{VanNoort2005}) and the CRISPRED data reduction pipeline \citep{Rodriguez2014}.
        We assume that the intensity calibration of the two lines is done as well as possible in the same way through the CRISPRED pipeline.
        Therefore, we are assuming that the relative intensities between the two lines are physically meaningful as assumed by our inversion technique.
        %\lfq{We need to say something about intensity calibration, particularly between the two channels, since the relative intensity of the line profiles is important. How is intensity calibration done, and are there any numbers quoted about relative or absolute photometric calibration?} 
        This event was previously analysed by \cite{2015Kuridze}, who presented the time-evolution of the H$\alpha$ and Ca\textsc{ii} 8542~\AA\ lines in small flaring regions, and compared these with RADYN forward modeling, driven by an electron beam with properties deduced from observed hard X-ray spectrum, commenting primarily on the relationship between plasma flows and line asymmetries.
                
        % Figure \ref{fig:images} shows the wing and core images of H$\alpha$ and Ca\textsc{ii} at $\sim$16:56~UTC,  just after the onset of the flare at $\sim$16:54~UTC.
        % The H$\alpha$ images in Fig. \ref{fig:images} clearly show the two flare ribbons with the structure looking sharper in the wing image (second column, top panel).
        % Only one of the ribbons is visible in the wing of Ca\textsc{ii} (first column, top panel) but both are clear in the core (first column, middle panel). \lfq{ I can see the ribbons in the Ca wing. They're faint, but they are present.}
        % We chose a region of interest of size 2$\times$2$\arcsec$ centred at the footpoint of the northern loop. \lf{A zoom in of this region is shown in the right-hand columns of Fig. \ref{fig:images}. }
        % The spectral line profiles from \lf{two pixels in} these observations are \lf{extracted, normalised to the maximum value in both lines (is this correct??)} and interpolated to the RADYN grid). These are shown in Fig. \ref{fig:lines}.
        
\begin{figure}[htp]
            \centering
            \includegraphics[width=0.49\textwidth]{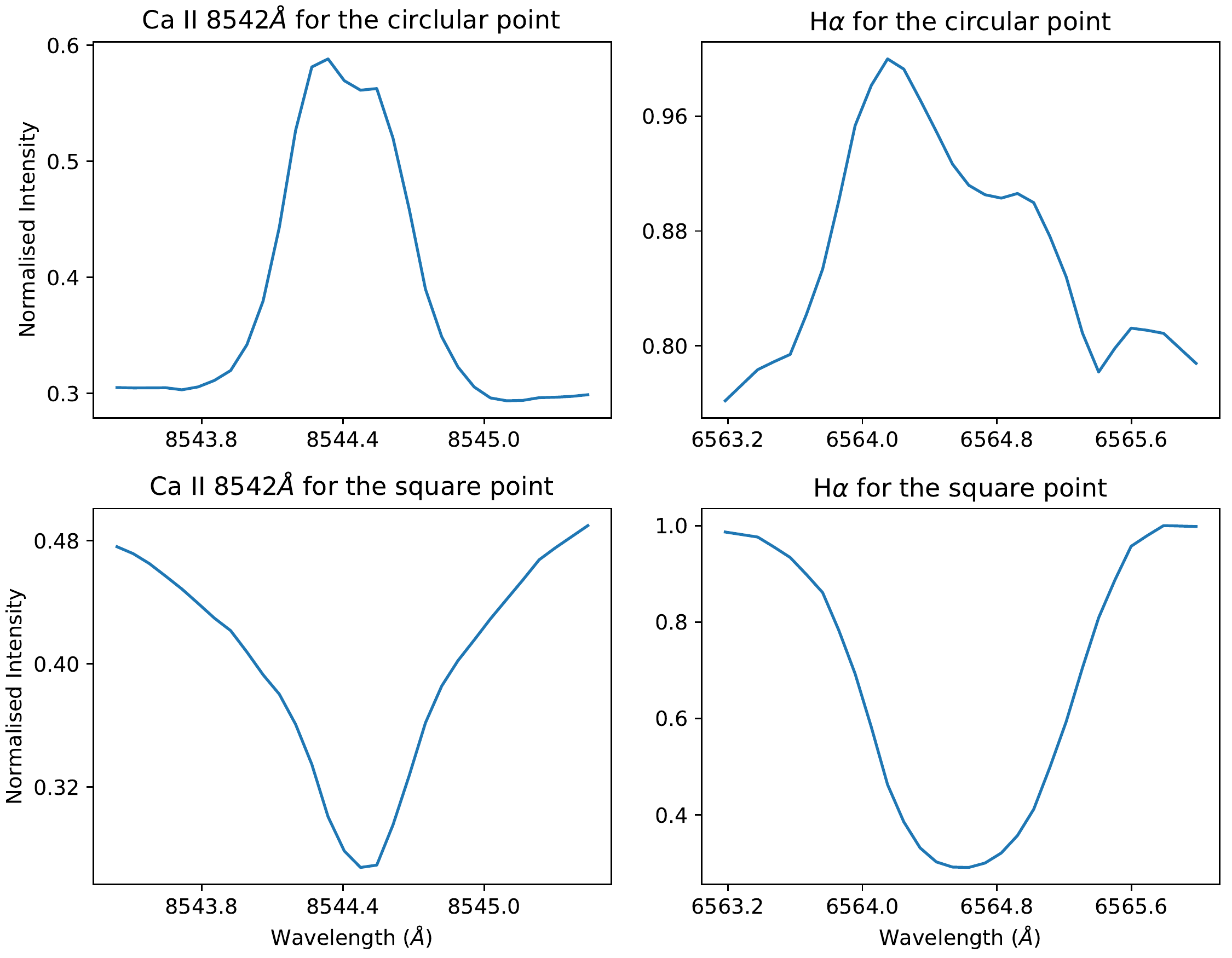}
            \caption{The spectral lines in Ca\textsc{ii} 8542\AA{} and H$\alpha$ for the two points selected in the region of interest. The top row shows one point on the flare ribbon and the bottom row shows one point off the flare ribbon. We perform inversions on both of these pairs of spectral lines.}
            \label{fig:lines}
        \end{figure}

        Figure \ref{fig:images} shows the wing and core images of Ca\textsc{ii} and H$\alpha$ at $\sim$16:56~UTC just after the onset of the flare at $\sim$16:54~UTC.
        These images clearly show the presence of two flare ribbons during the time of the observation.
        We chose two pixels to invert: one on the flare ribbon and one off the flare ribbon.
        These are indicated in the panels of Fig. \ref{fig:images} by a circle and square respectively.
        The spectral line profiles from the two pixels are extracted, normalised to the maximum value of the two lines and interpolated to the RADYN grid.
        These are shown in Fig. \ref{fig:lines}.
        
        The lines in the top row of Fig. \ref{fig:lines} are from a point on the flare ribbon, and those in the bottom row from a point off the flare ribbon (the circle and squares points, respectively, in Figure~\ref{fig:images}).
        The Ca\textsc{ii} 8542~\AA~ line profile for the circular point is characteristic during a flare.
        %\lfq{ But only one of these line profiles comes from the flare ribbon, so why do they both have these flare-like properties? Also, why is the bottom Ca profile, from off the flare ribbon, more intense than on the flare ribbon?}
        It is fully in emission and the core is slightly blueshifted (with respect to the vacuum wavelength) by $\sim 3.51$~km$\,$s$^{-1}$ with a slight wing asymmetry.
        %possibly pointing to chromospheric condensation or net upflows where the line wings are formed.
        The H$\alpha$ profile is highly asymmetric with the blue peak of the central reversal being much higher in emission than the red peak.
       For the square point, both profiles are heavily in absorption (indicative of the quiet Sun).
       The Ca\textsc{ii} and H$\alpha$ cores are slightly redshifted here (by $\sim$1.26~km$\,$s$^{-1}$ and $\sim$2.18~km$\,$s$^{-1}$, respectively) and both profiles have some asymmetry between the wings.
    
        To calculate the asymmetries in the profiles, we use a technique similar to that described in \citet{Mein1997,DePontieu2009} and \citet{2015Kuridze}.
        \begin{align}
            I_{B} &= \int_{\lambda_{0B} - \delta \lambda}^{\lambda_{0B} + \delta \lambda} \text{I} (\lambda) ~\text{d}\lambda\\
            I_{R} &= \int_{\lambda_{0R} - \delta \lambda}^{\lambda_{0R} + \delta \lambda} \text{I} (\lambda) ~\text{d}\lambda
        \end{align}
        where $\lambda_{0B}$ and $\lambda_{0R}$ are the centre wavelengths of the blue and red wings respectively and $\delta \lambda$ is the width of the wing from its centre wavelength.
        The wings are defined as being the area of the line one standard deviation away from the calculated intensity-averaged line core.
        The intensity-averaged line core is calculated via
        \begin{equation}
            \lambda_{0} = \frac{\int \text{I} (\lambda) ~\lambda ~\text{d}\lambda}{\int \text{I} (\lambda) ~\text{d}\lambda}
        \end{equation}
        which leads to us calculating the variance of the profile
        \begin{equation}
            \sigma^{2} = \frac{\int \text{I} (\lambda) ~(\lambda - \lambda_{0})^{2} ~\text{d}\lambda}{\int \text{I} (\lambda) ~\text{d}\lambda}
        \end{equation}
        Then the end of the blue wing and the start of the red wing are defined by $\lambda_{0} - \sigma$ and $\lambda_{0} + \sigma$ respectively, allowing us to calculate the central wavelengths for the wings and the half-width of the wings (i.e. $\lambda_{0B},\lambda_{0R}$ and $\delta \lambda$).
        These values along with the intensity ratio of the wings I$_{B}$/I$_{R}$ are presented in Table \ref{table:ints}.
        \begin{table*}[]
        \centering
\begin{tabular}{lllllll}
                                                & $\lambda_{0}$ [\AA{}] & $\sigma$ [\AA{}] & $\lambda_{0B}$ [\AA{}] & $\lambda_{0R}$ [\AA{}] & $\delta \lambda$ [\AA{}] & I$_{B}$/I$_{R}$ \\ \cline{2-7} 
\multicolumn{1}{l|}{H$\alpha$ on ribbon}        & 6564.57                 & 0.78               & 6563.49                  & 6565.68                  & 0.31                       & 0.996           \\ \cline{1-1}
\multicolumn{1}{l|}{Ca$\textsc{ii}$ on ribbon}  & 8544.43                 & 0.52               & 8543.67                  & 8545.20                  & 0.24                       & 1.032           \\ \cline{1-1}
\multicolumn{1}{l|}{H$\alpha$ off ribbon}       & 6564.58                 & 0.93               & 6563.41                  & 6565.75                  & 0.23                       & 0.983           \\ \cline{1-1}
\multicolumn{1}{l|}{Ca$\textsc{ii}$ off ribbon} & 8544.43                 & 0.62               & 8543.63                  & 8545.25                  & 0.19                       & 0.982          
\end{tabular}
\caption{The results of calculating the intensity-average line core and line standard deviation from moments analysis and using these values to calculate the asymmetries in the observed lines from Fig. \ref{fig:lines}. $\lambda_{0B}$ and $\lambda_{0R}$ are the central wavelengths of the blue and red wings of the line, respectively. $\delta \lambda$ is the half-width of the wings and I$_{B}$/I$_{R}$ is the wing intensity ratio.}
\label{table:ints}
\end{table*}
        The off-ribbon profiles both have red asymmetries of $\sim$~1.8~\% for calcium and $\sim$~1.7~\% for H$\alpha$.
        This corresponds to small positive velocity gradients or downflows in the region where the wings of these lines are formed.
        The calcium profile on the ribbon has a $\sim$~3.2~\% blue asymmetry while the H$\alpha$ profile has a red asymmetry of $\sim$~0.4~\%.
        This corresponds to small negative velocity gradients or upflows in the region where the wings of calcium are formed.
        
       It has been shown that the spectral lines we are considering should be symmetric about the line core in a static atmosphere \citep{Canfield1984,Fang1993,Cheng2006}, implying that the velocity field in the flaring atmosphere is responsible for the observed asymmetries.
       This is likely linked to chromospheric evaporation \citep{Neupert1968,Fisher1985,Graham2015} and condensation \citep{Ichimoto1984,Wulser1989}, which are the bulk expansion flows that occur in the rapidly heated flare chromosphere.
       However, a mapping between the observed asymmetry and the flow direction is complicated by absorption and emission in the moving plasma.
       For example, a blue asymmetry, as is observed in the Ca\textsc{ii} line on the flare ribbon, could be due to emission from upflowing plasma, or absorption by downflowing plasma, as argued for this flare by \cite{2015Kuridze}.
       %found that the blue asymmetry could not be solely due to plasma upflows and the red asymmetry could not be solely due to plasma downflows. 
       % This is due to the processes of chromospheric evaporation \citep{Neupert1968,Fisher1985,Graham2015} and condensation \citep{Ichimoto1984,Wulser1989}.
        %\cite{Ding1996} show that the red and blue asymmetries may both depend on downflows depending on where in the chromosphere the downflows occur.
       %\cite{2015Kuridze} main result is that red asymmetries might not be associated with downflows at all.}
         %This is indicative of chromospheric evaporation, occurring since the increase in upflow will cause an increase in optical depth of the blue wing but could also be due to increased downflows.
       %We conclude that the H$\alpha$ profile on the ribbon does not have a noticeable central reversal due to increased emission along the ribbon.
        
\begin{figure*}[htp]
            \centering
            \includegraphics[height=0.36\textheight]{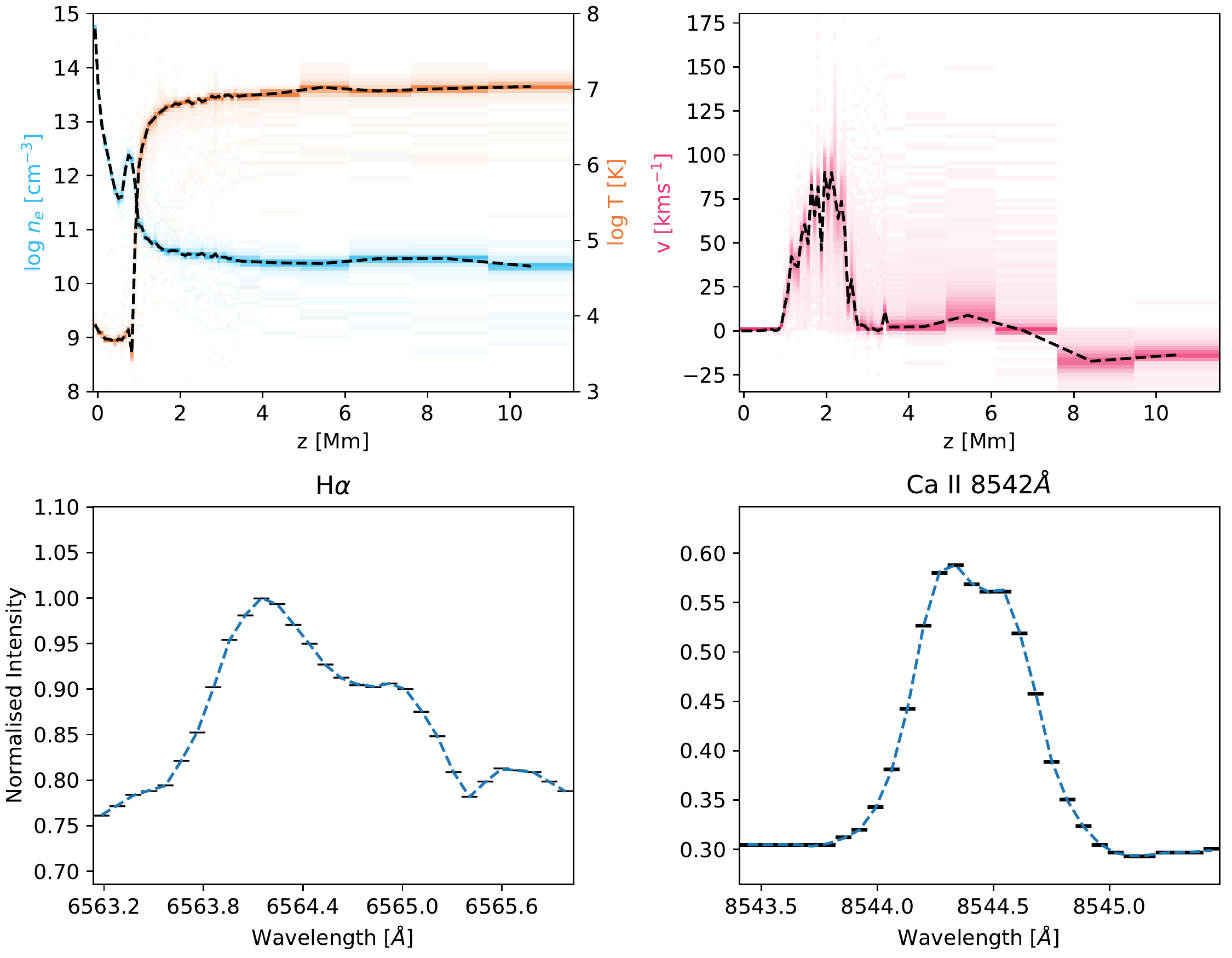}
            \caption{The inversion of the pixel on the flare ribbon. The top row shows the atmospheric parameters obtained from the inversion. The top left panel shows the electron density and temperature plotted on log scales and the top right panel shows the net velocity flow in our plasma. The plots were made by sampling the latent space 20000 times and plotting the results of the inversions as a 2-D histogram. The bins with the greatest density are the most likely values for the parameters at a certain height. The black dotted lines show the median profiles for each quantity. The bottom row shows the lines that were inverted. The blue dotted line in the bottom plots are the true line profiles. The black bins are the round trip generation of the spectral lines produced by performing the forward process on the sets of atmospheric parameters we obtain from the inversion.}
            \label{fig:inv_on_ribbon}
        \end{figure*}

\begin{figure*}[htp]
            \centering
            \includegraphics[height=0.36\textheight]{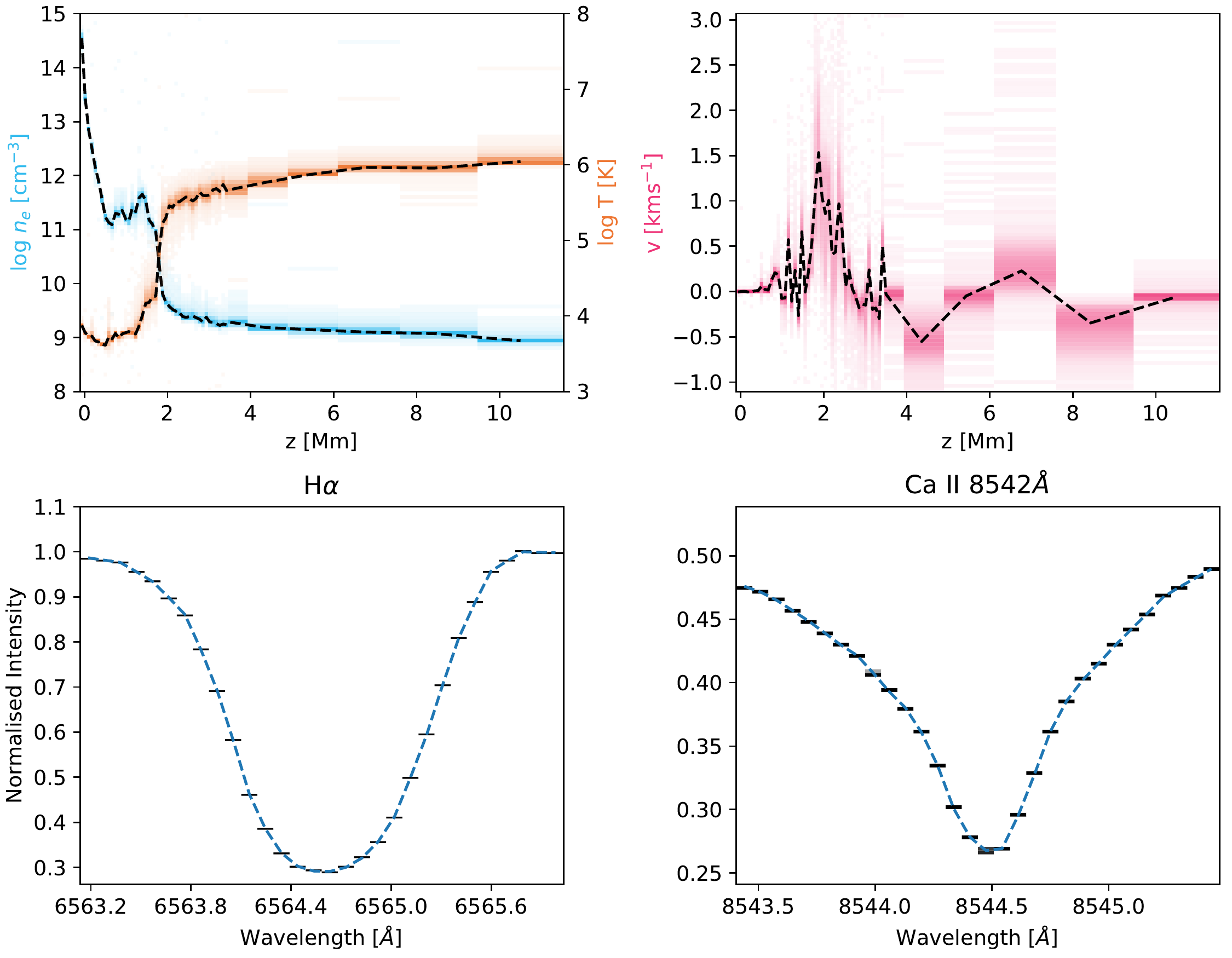}
            \caption{The inversion of the pixel off the flare ribbon. The plots have the same format as Fig. \ref{fig:inv_on_ribbon} and the latent space was also sampled for 20000 times.}
            \label{fig:inv_off_ribbon}
        \end{figure*}

        These observed spectral line profiles were propagated in the backwards direction through our INN (see Fig. \ref{fig:inn}) 20,000 times each with different random draws from the unit Gaussian latent space latent space (i.e. 20,000 inversions).
        The inversion of a single pixel takes $\sim$893~ms on an NVIDIA GTX~1050Ti and $\sim$84.5~s on an Intel Core i7-8700 CPU.
        The results of the inversions for the point on the flare ribbon are shown in Fig. \ref{fig:inv_on_ribbon} and for the point off the ribbon in Fig. \ref{fig:inv_off_ribbon}.
        As in the case of the model validation in Section~\ref{sec:validation}, the results are plotted as 2-D histograms
        (top rows of Fig. \ref{fig:inv_on_ribbon} \& \ref{fig:inv_off_ribbon}).
        %These bins are plotted using a power law normalisation to better show the outliers in our solutions.
        The dashed lines show the median profile for the parameters.
        This gives an approximation to the true solution from our inversion, as the median profile will pass through the bins with the highest densities.
        The bottom rows of these figures are plots of the observed spectral lines (dotted blue lines) and the densities of the round-trip profiles obtained by passing the results of the inversion back through the network in the forwards direction.
        This shows that each of the atmospheres we produce are viable for the production of these spectral lines with some curves being less likely due to the lack of density in the bins of the histogram (i.e. models with specific points in less dense bins are less likely to be the true solution). %\lfq{not clear what's meant here.. the lack of electron density? The lack of sample density in wavelength space? The fact that the 2D histograms aren't very concentrated in some cases?}
        
        %  \lfq{ In what follows you talk about `wings' without defining how far away in wavelength space from the nominal line centre you mean. It could mean an angstrom away - i.e. far wings -, but based on your argument I think you are still talking about less than half an angstrom from the nominal line centre? Better to specify the wavelength than to talk in generalities.}
                
        Examining the atmospheric profiles obtained from the inversions helps us interpret the line profiles generated. Looking first at the line asymmetries, we have previously remarked that for the on-ribbon pixel, the Ca\textsc{ii} line is slightly blueshifted with a blue asymmetry in the wings.
        According to \citet{Kerr2016}, the Ca\textsc{ii} 8542~\AA\ line during a flare is formed between 0.2 and 1.0~Mm above the photosphere, with the wings beyond $\pm$ 0.3~\AA\ from line centre formed between 0.2 and 0.4~Mm, i.e. in the upper photosphere/lower chromosphere.
        The line core within $\pm 0.3$~\AA~of line centre is formed above that.
        A steep positive velocity gradient in the area of core formation (0.9-1~Mm) explains the blueshifted core of our flare ribbon calcium profile.
        In the region of formation of the wings of this line, we observe a small positive upflow which would cause the observed blue asymmetry due to the emitting material moving upwards.
        %This suggests that the small blueshift is caused by increased absorption in a downward-moving (i.e. redshifted) chromospheric condensation, as in this region in our inverted velocity profile (0.2$\lesssim$z$\lesssim$1.2Mm) there is a negative velocity gradient. 
        % \lfq{I don't see this...looks pretty flat up to somewhat over 1Mm then there's an abrupt downward spike. So this wouldn't affect what I would call the line wings - but depends on our definition of wings.}
        % The H$\alpha$ profile for this pixel points to the chromospheric condensation solutions also.
        % \lf{This is consistent with the modeling of \cite{2015Kuridze}, which indicates that the H$\alpha$ profile forms below 1.2~Mm, with the wings beyond $\pm$ 0.5~\AA\ forming below 0.95~Mm and the core within $\pm$ 0.5~\AA\ forming above that height, such that increased absorption in the condensation will again lead to a blue asymmetry.}
        \citet{2015Kuridze} indicates that the H$\alpha$ profile forms below 1.2~Mm, with the wings forming below 0.95~Mm and the core forming above this height.
        The wings of the on-ribbon H$\alpha$ profile are very slightly asymmetric in favour of the red wing.
        In the region where the wings are formed, there is a small positive velocity gradient.
        This leads us to believe that there has been chromospheric evaporation in this region leading to an increase in optical depth in the region of the blue wing meaning that there will be more absorption in the blue wing.
        
        For our off-ribbon pixel, both profiles have small red asymmetries.
        This can be explained in our inverted atmosphere due to a turbulent flow where the lines are formed, which would also explain the asymmetries.
        Our velocity solution here is quite oscillatory.
        RADYN has an underlying 2~km$\,$s$^{-1}$ microturbulent velocity so the line profiles it produces are not as broad as those observed.
        Having learned that flows produce shifted emission, this oscillation is our model's attempt at making the lines the correct width.
        % This can be further explored by the inclusion of contribution functions in our \lf{future} analysis. \lfq{but since it's not a flare pixel should we expect either condensation or evaporation? May well be just small chromospheric flows, but then do we expect RADYN flare-like atmospheres to capture this?}
        
        The other main feature is the lack of a strong central reversal in H$\alpha$ during the flare. This is likely due to the source function being closer to the blackbody in the regions of line core formation in the flaring atmosphere compared to the non-flaring atmosphere. This may in turn be a result of the order of magnitude increase in the electron density at the line formation height in the flare, as indicated by the $n_e$ curves in Figures~\ref{fig:inv_on_ribbon} and~\ref{fig:inv_off_ribbon}.
        
        \section{Discussion and Conclusions} \label{sec:disc}
        %% Overall statement of what has been achieved
        We have presented a novel approach to obtaining the distribution of solar atmospheric properties $n_e, T$ and bulk flow speed $v$ from observed H$\alpha$ and Ca\textsc{ii} 8542~\AA\ spectral line profiles, using an invertible neural network trained on RADYN flare models. The network learns a bijective approximation to the forwards and inverse problems of mapping atmospheric snapshots to (observable) spectral line profiles and vice versa.
        Our initial results are very promising when tested on a flare previously analysed by \citet{2015Kuridze}, aligning well with the their results as discussed in Section~\ref{sec:inv}.
        
        %% Relates what has been achieved to previous work that readers should be more familiar with, and somewhat 'demystifies' the INN
        The INN method of atmospheric inversion represents a significant theoretical step forward in the field of inversion.
        Taking the process of training and applying the INN as a whole, it is comparable to the process performed by existing non local thermodynamic equilibrium inversion tools, which are typically composed of a forward model for computing the line profiles from an atmosphere such as RH \citep{Uitenbroek2001}, and an ``inversion engine'' that is responsible for determining the necessary perturbations to the atmosphere to produce a best-fit line profile.
        Our INN first learns the forward process from our training data, but due to the bijective nature of the mapping, a perturbative solution approach is not required, as all of the information lost in the forward process can be restored through the latent space.
        The models that take this ``inversion engine'' approach, such as STiC \citep{DelaCruzRodriguez2018} and NICOLE \citep{Socas-Navarro2014} are effectively performing a walk through the latent space guided by their ``inversion engines''.
        There is no guarantee of solution uniqueness from those approaches as the entire latent space is not visited.
        With the INN approach the useful extent of the latent space is learned during training, and it is therefore trivial to span the latent space with multiple draws of the unit multivariate normal distribution. 
        
        %%Contrasts INN with pattern-matching approaches
        As our INN was trained on RADYN data it is important to stress that it can only generate RADYN-like  solutions and this should be taken into account when analysing any atmospheric inversions performed. The RADYN training atmospheres also include the specific assumption of heating and non-thermal excitations by an electron beam from the corona.
        As a counterpoint to this, it is important to note that the INN does not simply ingest the grid of RADYN simulations and return a closely matched or interpolated template (an approach  used for example by \cite{Beck2015} in the the fast inversion of Ca\textsc{ii} 8542~\AA\ spectropolarimetric data.)
        Instead, the INN has learned a bijective mapping between the input space containing the atmospheric parameters and the output space containing the line profiles and the explicit latent space. In the inverse process the line profiles are complemented by the latent space to remove ambiguities due to information lost in the forward process.
        The model's validation on the unseen testing set should ensure that the atmospheres recovered are physically reasonable, and that the model has learnt to relate the emergent line profiles with properties of the atmosphere.
        
        %% describes the advantages of the INN
        The INN method is fast, as it ``front-loads'' a large portion of the computational work, by requiring a large training set in the form of RADYN simulations followed by approximately 1 day of training on an NVIDIA GTX~1050Ti GPU. The result of this precomputation is that inference is then extremely rapid, while still drawing on a very complex physical model. % than the one used in Caisar \citep{Beck2015}. \lfq{non sequitur. Caisar has not been mentioned before, nor the type of model it uses.} 
        The complex model is needed for the flare problem, where assumptions of hydrostatic and local thermodynamic equilibrium cannot hold, and steep gradients are expected to form. This presents a further advantage of the INN method for flares, since to reduce the size of the parameter space and allow an ``inversion engine'' to converge in a reasonable amount of time, all other inversion codes currently assume that the atmosphere is in hydrostatic equilibrium \citep{DelaCruzRodriguez2018,Socas-Navarro2014} and use $<$10 nodes in the atmosphere where the parameters are computed with various interpolation techniques used between these. 
        
        %%expands on the INN's capacity to deal with non-LTE 
        As found in \cite{Brown2018} the non-equilibrium level population and ionisation effects present in RADYN, including those due to direct excitations by non-thermal electrons, cause significant deviations between the line profiles computed with these populations and those computed under the assumption of statistical equilibrium in RH \citep{Uitenbroek2001}. 
        %In flaring cases it appears that the differences due to non-equilibrium effects are often more significant than those due the the differences between complete and partial redistribution.
        Because our model is trained on RADYN data, the associated line profiles are based on RADYN's non-equilibrium formalism, and its assumption of complete redistribution (i.e. the frequency of an absorbed photon that leads to an excited state and that of the resulting emitted photon are assumed to be independent).
        These effects are therefore learned by the INN. 
        It is interesting that, even with limited atmospheric infomation, i.e. $n_e, T$, and $v$, which are a far from complete description of the state of the atmosphere, the INN was nevertheless able to very successfully reproduce the emission from the unseen RADYN snapshots from the F-CHROMA grid. This implies that sufficient non-LTE and non-hydrostatic equilibrium information about local `microscopic' (ionisation, level populations), `macroscopic' (gas pressure, opacity), and non-local physics (conduction, radiative backwarming) must be encoded in these three parameters and their variation through the atmosphere.
        
        %% problem with our results
        Inversions of pixels on the flare ribbon performed in Sec.~\ref{sec:inv}, suggest significant oscillations in the velocity profile in the transition region (e.g. Fig.~\ref{fig:inv_on_ribbon}).
        These oscillations do not simply appear on the median line, but appear with a similar form on many of the individual velocity profiles obtained from the inversion.
        This may in part be due to RADYN using a conservative 2~km$\,$s$^{-1}$ microturbulent velocity throughout the atmosphere.
        Studies with the Interface Region Imaging Spectrograph (IRIS; \cite{Title2014}) have required significantly higher values to explain the non-thermal broadening in Mg\textsc{ii} h \& k in chromospheric plage. \cite{Carlsson2015} find a value $\sim$7~km$\,$s$^{-1}$) and the inversions performed with STiC \citep{DelaCruzRodriguez2018} suggest a value $\sim$8~km$\,$s$^{-1}$ for the same observation.
        We suggest then that the INN needs to broaden the line to match observations and uses an oscillating velocity, and higher temperature, in the $\tau=1$ region to achieve this. 
        To better constrain the parameters in the upper chromosphere and transition region requires computation of lines such as Mg\textsc{ii}~h \& k, or Si\textsc{IV}~1403~\AA\, but these are currently not calculated in RADYN.
        Whilst the emission from Mg\textsc{ii}~h \& k could be computed from populations in statistical equilibrium using RH it is essential to verify whether the non-equilibrium effects are important for these lines in flares.

        There are several additional assumptions made during the training process that need to be considered when applying the INN.
        \begin{enumerate}
        \item Only the line profiles from the $\mu\approx0.9531$ ray angle were included in the training set.
        This is the emergent radiation at an angle $\cos^{-1}{\mu}\approx17.6^\circ$ to the normal of the atmospheric layers of the plane parallel atmosphere used in RADYN.
        The emergent radiation detected from the flare discussed in Sec.~\ref{sec:inv} is approximately 37$^\circ$ from the local vertical.
        Assuming a plane parallel atmosphere, the layers appear thicker by a factor of $1/\mu$ than their depth along the normal to the atmosphere, so shallower layers may have a more significant effect than is predicted by the training set.
        The altitude stratification in the training set is perpendicular to the solar surface at this assumed $\mu$ ray angle to the observer.
        \item Although different beam parameters are used, the simulations in the F-CHROMA RADYN grid all use the same 20s triangular heating pulse, leading to a particular temporal sequence in the run of atmospheric properties that may not occur for different heating profiles (or indeed for different heating methods).
        As the inversions performed in Sec.~\ref{sec:inv} appear well-constrained, this does not appear to be an issue. 
        \end{enumerate}
         
        %As RADYN is a radiation-hydrodynamic model no magnetic effects are considered and the inversions cannot yield information on the magnetic field.
        
        \vspace{1em}
        To summarise, our novel technique using an invertible neural network trained with simulations from the radiation-hydrodynamics model RADYN to solve the inverse problem of determining the solar atmospheric parameters given chromospheric spectral line profiles, lifts several restrictions that affect other inversion methods, such as enforcing hydrostatic equilibrium, that make these methods unusable for energetic atmospheres.
        The method is fast to train, very rapid to apply to data, has proven accurate on unseen validation tests, and early results are very convincing and in broad agreement with previous analyses.
        This method of solving inverse problems is computationally tractable when a prior forward exists and could be leveraged to solve many other astrophysical problems.
        The code is available online under the MIT license\footnote{\url{https://opensource.org/licenses/MIT}} at \url{https://github.com/Goobley/Radynversion} and will soon be added to the RadynPy\footnote{\url{https://github.com/Goobley/radynpy}} \citep{RadynPy} python package.
        
        \section*{Acknowledgements}
        CMJO acknowledges support from the UK's Science and Technology Facilities Council (STFC) doctoral training grant ST/R504750/1.
        JAA acknowledges a data-intensive science studentship with the STFC `ScotDIST' centre for doctoral training supported by grant ST/R504750/1. 
        LF acknowledges support from STFC grant ST/P000533/1. The authors are grateful to M. Carlsson and the F-CHROMA collaboration for the production and availability of the grid of RADYN simulations. The research leading to these results has received funding from the European Communityʼs Seventh Framework Programme (FP7/2007-2013) under grant agreement No. 606862 (F-CHROMA), and from the Research Council of Norway through the Programme for Supercomputing.
        The authors would like to thank P.J.A. Sim\~{o}es for helpful discussions and general constructive advice.
        The authors are also grateful to the reviewer for helpful comments and corrections.
        
        \appendix
       
        \section{Maximum Mean Discrepancy}\label{App:Mmd}
        This following section draws heavily on \cite{Gretton2012} and the lectures on this topic given at the Machine Learning Summer School Madrid 2018\footnote{available at \url{http://www.gatsby.ucl.ac.uk/~gretton/teaching.html}}.
        
        Training the INN is made possible by the use of the Maximum Mean Discrepancy (MMD).
        The MMD is a technique for determining the distance between probability distributions $P$ and $Q$ using observations $X=\{x_1,\ldots,x_m\}$ and $Y=\{y_1,\ldots,y_n\}$ drawn in an independent and identically distributed fashion from $P$ and $Q$ respectively. 
        The MMD can be mathematically expressed as
        \begin{equation}
        \begin{aligned}
        \label{Eq:Mmd}
        \textrm{MMD}^2 &= ||\mu_P-\mu_Q||_\mathcal{F}^2\\
        &= \langle\mu_P,\,\mu_P\rangle_\mathcal{F} + \langle\mu_Q,\,\mu_Q\rangle_\mathcal{F} - 2\langle\mu_P,\,\mu_Q\rangle_\mathcal{F}
        \end{aligned}
        \end{equation}
        where
        $\mathcal{F}$ is a Reproducing Kernel Hilbert Space (RKHS) known as the feature space, with elements known as features,
        $\langle\cdot\,,\,\cdot\rangle_\mathcal{F}$ denotes the inner product in the feature space,
        and $\mu_A$ represents the expectation vector of the features of $\mathcal{F}$ evaluated for the distribution $A$. 
        
        Let $X$ be a non-empty space with positive definite kernel $k : X\times X \rightarrow \mathbb{R}$ and $\phi : X \rightarrow \mathcal{F}$ a feature map, then for all $x,\,y \in X$
        \begin{equation}
            \label{Eq:KernelProp}
            k(x,y) = \langle\phi(x), \phi(y)\rangle_\mathcal{F}
        \end{equation}
        
        The features spaces of kernels such as the Gaussian kernel 
        
        \begin{equation*}
        k(x,y) = e^\frac{||x-y||^2}{2\sigma^2},\quad\sigma>0
        \end{equation*}
        
        are in fact infinite dimensional but the kernel trick of \eqref{Eq:KernelProp} allows the inner product between vectors in this space to be written in closed form.
        The reproducing property of the RKHS states simply that under the inner product of features in $\mathcal{F}$ the kernel will always be recovered.
        For a positive definite kernel there is a unique RKHS $\mathcal{F}$ with reproducing kernel $k$, whose features are a subset of $\mathcal{F}$, therefore a feature map is not unique, but the kernel is.
        
        $\mu_P$ from \eqref{Eq:Mmd} can then be written in terms of the features of $\mathcal{F}$
        
        \begin{equation}
            \mu_P=[\ldots\,\mathbb{E}_P[\phi_i(X)]\,\ldots]
        \end{equation}
        
        where $\mathbb{E}_P$ denotes the expectation value of its argument with respect to $P$ and $\phi_i$ is the $i$-th feature of $\phi$.
        From this definition we can write
        
        \begin{equation}
            \langle\mu_P,\,\mu_Q\rangle_\mathcal{F} = \mathbb{E}_{P,Q}[k(x,y)]
        \end{equation}
        
        where $\mathbb{E}_{P,Q}[k(\cdot\,,\,\cdot)]$ denotes the expected kernel of $P$ and $Q$ where $x\sim P$ and $y\sim Q$, (and $a \sim A$ indicates that $a$ is drawn in an unbiased way from $A$).
        
        Now, from the expansion in \eqref{Eq:Mmd} we have
        
        \begin{equation}
        \begin{aligned}
            \textrm{MMD}^2 &= ||\mu_P-\mu_Q||^2_\mathcal{F} \\
            &= \mathbb{E}_P[k(x,x^\prime)] + \mathbb{E}_Q[k(y,y^\prime)] - 2 \mathbb{E}_{P,Q}[k(x,y)].
        \end{aligned}
        \end{equation}
        
        For finite observations $X$ and $Y$ (of length $n$) this then gives an unbiased sample estimate of the MMD
        
        \begin{equation}
            \widehat{\textrm{MMD}}_u^2 = \frac{1}{n(n-1)}\sum_{i\neq j} k(x_i, x_j) + \frac{1}{n(n-1)}\sum_{i\neq j} k(y_i, y_j) - \frac{2}{n^2}\sum_{i,\,j}k(x_i, y_j).
        \end{equation}
        
        Due to the efficiency of matrix operations used to compute the MMD loss in our training scheme we compute a biased sample estimate of the MMD
    
        \begin{equation}
            \widehat{\textrm{MMD}}_b^2 = \frac{1}{n^2}\sum_{i,\,j}\left(k(x_i, x_j) + k(y_i, y_j) - 2k(x_i, y_i)\right).
        \end{equation}
        
        The bias on this statistic simply increases the expected MMD result, but has the advantage of remaining positive even when $P=Q$, which works better with the optimiser used to train the INN.
        
        \bibliographystyle{apalike}
        \bibliography{ref.bib}

\begin{thebibliography}{}

\bibitem[Allred et~al., 2005]{Allred2005}
Allred, J., Hawley, S.~L., Abbett, W., and Carlsson, M. (2005).
\newblock {Radiative Hydrodynamics Models of the Optical and Ultraviolet
  Emission from Solar Flares}.
\newblock {\em The Astrophysical Journal}, 630:573--586.

\bibitem[Allred et~al., 2015]{Allred2015}
Allred, J.~C., Kowalski, A.~F., and Carlsson, M. (2015).
\newblock {A Unified Computational Model for Solar and Stellar Flares}.
\newblock {\em Astrophysical Journal}, 809(1):104.

\bibitem[Ardizzone et~al., 2018]{Ardizzone2018}
Ardizzone, L., Kruse, J., Wirkert, S., Rahner, D., Pellegrini, E.~W., Klessen,
  R.~S., Maier-hein, L., Rother, C., and K{\"{o}}the, U. (2018).
\newblock {Analyzing Inverse Problems with Invertible Neural Networks}.
\newblock {\em ArXiv e-prints}, pages 1--18.

\bibitem[Asensio~Ramos et~al., 2008]{Ramos2008}
Asensio~Ramos, A., Bueno, J.~T., and Degl'Innocenti, E.~L. (2008).
\newblock {Advanced Forward Modeling and Inversion of Stokes Profiles Resulting
  from the Joint Action of the Hanle and Zeeman Effects}.
\newblock {\em Astrophysical Journal}, 683:542--565.

\bibitem[Beck et~al., 2015]{Beck2015}
Beck, C., Choudhary, D.~P., Rezaei, R., and Louis, R.~E. (2015).
\newblock {FAST INVERSION OF SOLAR Ca II SPECTRA}.
\newblock {\em The Astrophysical Journal}, 798(2):100--108.

\bibitem[Bradshaw and Cargill, 2013]{Bradshaw2013}
Bradshaw, S.~J. and Cargill, P.~J. (2013).
\newblock {The influence of numerical resolution on coronal density in
  hydrodynamic models of impulsive heating}.
\newblock {\em Astrophysical Journal}, 770(1).

\bibitem[Brown et~al., 2018]{Brown2018}
Brown, S.~A., Fletcher, L., Kerr, G.~S., Labrosse, N., and Kowalski, A.~F.
  (2018).
\newblock {Modelling of the hydrogen lyman lines in solar flares}.
\newblock {\em The Astrophysical Journal}.

\bibitem[Canfield et~al., 1984]{Canfield1984}
Canfield, R.~C., Gunkler, T.~A., and Ricchiazzi, P.~J. (1984).
\newblock {The H$\alpha$ Spectral Signatures of Solar Flare Nonthermal
  Electrons, Conductive Flux, and Coronal Pressure}.
\newblock {\em The Astrophysical Journal}, 282:296--307.

\bibitem[Carlsson et~al., 2015]{Carlsson2015}
Carlsson, M., Leenaarts, J., and {De Pontieu}, B. (2015).
\newblock {WHAT DO IRIS OBSERVATIONS of Mg II k TELL US about the SOLAR PLAGE
  CHROMOSPHERE?}
\newblock {\em Astrophysical Journal Letters}, 809(2):L30.

\bibitem[Carlsson and Stein, 1992]{Carlsson1992}
Carlsson, M. and Stein, R. (1992).
\newblock {Non-LTE Radiating Acoustic Shocks and Ca II K2V Bright Points}.
\newblock {\em The Astrophysical Journal}, 397.

\bibitem[Carlsson and Stein, 1997]{Carlsson1997}
Carlsson, M. and Stein, R.~F. (1997).
\newblock {Formation of Solar Calcium H and K Bright Grains}.
\newblock {\em The Astrophysical Journal}, 481(1):500--514.

\bibitem[Cauzzi et~al., 2014]{2014Cauzzi}
Cauzzi, G., Fletcher, L., Mathioudakis, M., Carlsson, M., Heinzel, P.,
  Berlicki, A., and Zuccarello, F. (2014).
\newblock {F-CHROMA.Flare Chromospheres: Observations, Models and Archives}.
\newblock In {\em American Astronomical Society Meeting Abstracts {\#}224},
  volume 224 of {\em American Astronomical Society Meeting Abstracts}, page
  123.39.

\bibitem[Cheng et~al., 2006]{Cheng2006}
Cheng, J.~X., Ding, M.~D., and Li, J.~P. (2006).
\newblock {Diagnostics of the heating processes in solar flares using
  chromospheric spectral lines}.
\newblock {\em Astrophysical Journal}, 653(1 I):733--738.

\bibitem[Cybenko, 1989]{Cybenko1989}
Cybenko, G. (1989).
\newblock {Approximation by Superpositions of a Sigmoidal Function}.
\newblock {\em Mathematics of Control, Signals, and Systems}, 2:303--314.

\bibitem[da~Costa et~al., 2016]{Costa2016}
da~Costa, F.~R., Kleint, L., Petrosian, V., Liu, W., and Allred, J.~C. (2016).
\newblock {Data-Driven Radiative Hydrodynamic Modeling of the 2014 March 29
  X1.0 Solar Flare}.
\newblock {\em The Astrophysical Journal}, 827(1):38.

\bibitem[{de la Cruz Rodriguez} et~al., 2018]{DelaCruzRodriguez2018}
{de la Cruz Rodriguez}, J., Leenaarts, J., Danilovic, S., and Uitenbroek, H.
  (2018).
\newblock {STiC – A multi-atom non-LTE PRD inversion code for full-Stokes
  solar observations}.
\newblock {\em Astronomy and Astrophysics}, pages 1--13.

\bibitem[{de la Cruz Rodr{\'\i}guez} et~al., 2015]{Rodriguez2014}
{de la Cruz Rodr{\'\i}guez}, J., {L{\"o}fdahl}, M.~G., {S{\"u}tterlin}, P.,
  {Hillberg}, T., and {Rouppe van der Voort}, L. (2015).
\newblock {CRISPRED: A data pipeline for the CRISP imaging spectropolarimeter}.
\newblock {\em \aap}, 573:A40.

\bibitem[{De Pontieu} et~al., 2009]{DePontieu2009}
{De Pontieu}, B., McIntosh, S.~W., Hansteen, V.~H., and Schrijver, C.~J.
  (2009).
\newblock {Observing the roots of solar coronal heating - In the chromosphere}.
\newblock {\em ASTROPHYSICAL JOURNAL LETTERS}, 701(1 PART 2).

\bibitem[{De Pontieu} et~al., 2014]{Title2014}
{De Pontieu}, B., {Title}, A.~M., {Lemen}, J.~R., {Kushner}, G.~D., {Akin},
  D.~J., {Allard}, B., {Berger}, T., {Boerner}, P., {Cheung}, M., {Chou}, C.,
  {Drake}, J.~F., {Duncan}, D.~W., {Freeland}, S., {Heyman}, G.~F., {Hoffman},
  C., {Hurlburt}, N.~E., {Lindgren}, R.~W., {Mathur}, D., {Rehse}, R.,
  {Sabolish}, D., {Seguin}, R., {Schrijver}, C.~J., {Tarbell}, T.~D.,
  {W{\"u}lser}, J.-P., {Wolfson}, C.~J., {Yanari}, C., {Mudge}, J.,
  {Nguyen-Phuc}, N., {Timmons}, R., {van Bezooijen}, R., {Weingrod}, I.,
  {Brookner}, R., {Butcher}, G., {Dougherty}, B., {Eder}, J., {Knagenhjelm},
  V., {Larsen}, S., {Mansir}, D., {Phan}, L., {Boyle}, P., {Cheimets}, P.~N.,
  {DeLuca}, E.~E., {Golub}, L., {Gates}, R., {Hertz}, E., {McKillop}, S.,
  {Park}, S., {Perry}, T., {Podgorski}, W.~A., {Reeves}, K., {Saar}, S.,
  {Testa}, P., {Tian}, H., {Weber}, M., {Dunn}, C., {Eccles}, S., {Jaeggli},
  S.~A., {Kankelborg}, C.~C., {Mashburn}, K., {Pust}, N., {Springer}, L.,
  {Carvalho}, R., {Kleint}, L., {Marmie}, J., {Mazmanian}, E., {Pereira},
  T.~M.~D., {Sawyer}, S., {Strong}, J., {Worden}, S.~P., {Carlsson}, M.,
  {Hansteen}, V.~H., {Leenaarts}, J., {Wiesmann}, M., {Aloise}, J., {Chu},
  K.-C., {Bush}, R.~I., {Scherrer}, P.~H., {Brekke}, P., {Martinez-Sykora}, J.,
  {Lites}, B.~W., {McIntosh}, S.~W., {Uitenbroek}, H., {Okamoto}, T.~J.,
  {Gummin}, M.~A., {Auker}, G., {Jerram}, P., {Pool}, P., and {Waltham}, N.
  (2014).
\newblock {The Interface Region Imaging Spectrograph (IRIS)}.
\newblock {\em \solphys}, 289:2733--2779.

\bibitem[Dinh et~al., 2014]{Dinh2015}
Dinh, L., Krueger, D., and Bengio, Y. (2014).
\newblock {NICE : Non-Linear Independent Components Estimation}.
\newblock {\em ArXiv e-prints}, pages 1--13.

\bibitem[Dinh et~al., 2016]{Dinh2017}
Dinh, L., Sohl-Dickstein, J., and Bengio, S. (2016).
\newblock {Density Estimation using Real NVP}.
\newblock {\em ArXiv e-prints}.

\bibitem[Dorfi and Drury, 1987]{Dorfi1987}
Dorfi, E.~A. and Drury, L.~O. (1987).
\newblock {Simple adaptive grids for 1-D initial value problems}.
\newblock {\em Journal of Computational Physics}, 69(1):175--195.

\bibitem[Fang et~al., 1993]{Fang1993}
Fang, C., Henoux, J., and Gan, W. (1993).
\newblock {Diagnostics of Non-thermal Processes in Chromospheric Flares}.
\newblock {\em Astronomy {\&} Astrophysics}, 274:917--922.

\bibitem[Fisher et~al., 1985]{Fisher1985}
Fisher, G.~H., Canfield, R.~C., and Mcclymont, A.~N. (1985).
\newblock {Flare Loop Radiative Hydrodynamics. V. Response to Thick-Target
  Heating}.
\newblock {\em The Astrophysical Journal}, 289:414--424.

\bibitem[Fletcher et~al., 2007]{Fletcher2007}
Fletcher, L., Hannah, I.~G., Hudson, H.~S., and Metcalf, T.~R. (2007).
\newblock {A TRACE White Light and RHESSI Hard X-Ray Study of Flare
  Energetics}.
\newblock {\em Astrophysical Journal}, 656:1187--1196.

\bibitem[Graham and Cauzzi, 2015]{Graham2015}
Graham, D.~R. and Cauzzi, G. (2015).
\newblock {TEMPORAL EVOLUTION of MULTIPLE EVAPORATING RIBBON SOURCES in A SOLAR
  FLARE}.
\newblock {\em Astrophysical Journal Letters}, 807(2):L22.

\bibitem[Gretton et~al., 2012]{Gretton2012}
Gretton, A., Borgwardt, K., Rasch, M., Sch\"{o}lkopf, B., and Smola, A. (2012).
\newblock {A Kernel Two-Sample Test}.
\newblock {\em Journal of Machine Learning Research}, 13:723--773.

\bibitem[Heinzel et~al., 2015]{Heinzel2015}
Heinzel, P., Ka{\v{s}}parov{\'{a}}, J., Varady, M., Karlick{\'{y}}, M., and
  Moravec, Z. (2015).
\newblock {Numerical RHD simulations of flaring chromosphere with Flarix}.
\newblock {\em Proceedings of the International Astronomical Union},
  11(S320):233--238.

\bibitem[Ichimoto and Kurokawa, 1984]{Ichimoto1984}
Ichimoto, K. and Kurokawa, H. (1984).
\newblock {H$\alpha$ Red Asymmetry of Solar Flares}.
\newblock {\em Solar Physics}, 93(105).

\bibitem[Kennedy et~al., 2015]{Kennedy2015}
Kennedy, M.~B., Milligan, R.~O., Allred, J.~C., Mathioudakis, M., and Keenan,
  F.~P. (2015).
\newblock {Radiative hydrodynamic modelling and observations of the X-class
  solar flare on 2011 March 9}.
\newblock {\em Astronomy {\&} Astrophysics}, 72:1--12.

\bibitem[Kerr et~al., 2016]{Kerr2016}
Kerr, G.~S., Fletcher, L., Russell, A. J.~B., and Allred, J.~C. (2016).
\newblock {Simulations of the Mg II K and Ca II 8542 Lines from an Alfv{\'{e}}n
  Wave-Heated Flare Chromosphere}.
\newblock {\em The Astrophysical Journal}, 827(2):1--16.

\bibitem[Kingma and Ba, 2014]{Kingma2014}
Kingma, D.~P. and Ba, J.~L. (2014).
\newblock {Adam: A method for stochastic optimization}.
\newblock {\em 3rd International Conference for Learning Representations}.

\bibitem[Kowalski et~al., 2017]{Kowalski2017}
Kowalski, A.~F., Allred, J.~C., Daw, A., Cauzzi, G., and Carlsson, M. (2017).
\newblock {The Atmospheric Response to High Nonthermal Electron Beam Fluxes in
  Solar Flares. I. Modeling the Brightest NUV Footpoints in the X1 Solar Flare
  of 2014 March 29}.
\newblock {\em The Astrophysical Journal}, 836(1):12.

\bibitem[Kretzschmar, 2011]{Kretzschmar2011}
Kretzschmar, M. (2011).
\newblock {The Sun as a star : observations of white-light flares}.
\newblock {\em Astronomy {\&} Astrophysics}, 84:1--7.

\bibitem[Krucker et~al., 2011]{Krucker2011}
Krucker, S., Hudson, H.~S., Jeffrey, N.~L., Battaglia, M., Kontar, E.~P., Benz,
  A.~O., Csillaghy, A., and Lin, R.~P. (2011).
\newblock {High-resolution imaging of solar flare ribbons and its implication
  on the thick-target beam model}.
\newblock {\em Astrophysical Journal}, 739(2).

\bibitem[Kuridze et~al., 2017]{Kuridze2017}
Kuridze, D., Henriques, V., Mathioudakis, M., Koza, J., Zaqarashvili, T.~V.,
  Ryb{\'{a}}k, J., Hanslmeier, A., and Keenan, F.~P. (2017).
\newblock {Spectroscopic Inversions of the Ca II 8542 Å Line in a C-class
  Solar Flare}.
\newblock {\em The Astrophysical Journal}, 846(1):9.

\bibitem[Kuridze et~al., 2018]{Kuridze2018}
Kuridze, D., Henriques, V., Mathioudakis, M., van~der Voort, L.~R.,
  Rodr{\'{i}}guez, J. d. l.~C., and Carlsson, M. (2018).
\newblock {Spectropolarimetric inversions of the Ca II 8542 \AA~ line in a
  M-class solar flare}.
\newblock {\em The Astrophysical Journal}, 860(1):10.

\bibitem[Kuridze et~al., 2015]{2015Kuridze}
Kuridze, D., Mathioudakis, M., Sim{\~{o}}es, P., {Rouppe van der Voort}, L.,
  Carlsson, M., Jafarzadeh, S., Allred, J., Kowalski, A., Kennedy, M.,
  Fletcher, L., Graham, D., and Keenan, F. (2015).
\newblock {H$\alpha$ Line Profile Asymmetries and the Chromospheric Flare
  Velocity Field}.
\newblock {\em Astrophysical Journal}, 813:125.

\bibitem[Mein et~al., 1997]{Mein1997}
Mein, P., Mein, N., Heinzel, P., Kneer, F., Uexkull, M. V. O.~N., and Staiger,
  J. (1997).
\newblock {Flare multi-line 2d-spectroscopy}.
\newblock {\em Solar Physics}, 172(161):161--170.

\bibitem[Metcalf et~al., 1990]{Metcalf1990}
Metcalf, T.~R., Canfield, R.~C., Avrett, E.~H., and Metcalf, F.~T. (1990).
\newblock {Flare Heating and Ionization of the Low Solar Chromosphere. I.
  Inversion Methods for Mg I 4571 and 5173}.
\newblock {\em The Astrophysical Journal}, 350:463--474.

\bibitem[Milligan et~al., 2014]{Milligan2014}
Milligan, R.~O., Kerr, G.~S., Dennis, B.~R., Hudson, H.~S., Fletcher, L.,
  Allred, J.~C., Chamberlin, P.~C., Ireland, J., Mathioudakis, M., and Keenan,
  F.~P. (2014).
\newblock {The radiated energy budget of chromospheric plasma in a major solar
  flare deduced from multi-wavelength observations}.
\newblock {\em Astrophysical Journal}, 793(2).

\bibitem[Neupert, 1968]{Neupert1968}
Neupert, W. (1968).
\newblock {Comparison of Solar X-ray Line Emission with Microwave Emission
  During Flares}.
\newblock {\em The Astrophysical Journal}, 153:3.

\bibitem[Osborne, 2019]{RadynPy}
Osborne, C. M.~J. (2019).
\newblock Goobley/radynpy: Contribution function update.

\bibitem[Raschka, 2015]{raschka_2015}
Raschka, P. i. M.~S. (2015).
\newblock {\em {Python Machine Learning}}.
\newblock Packt Publishing Limited.

\bibitem[Rumelhart et~al., 1986]{Rumelhart1986}
Rumelhart, D., Hinton, G., and Williams, R. (1986).
\newblock {Learning Internal Representations by Error Propagation}.
\newblock In {\em Parallel distributed processing: explorations in the
  microstructure of cognition}, volume~1, pages 318--362.

\bibitem[Scharmer, 2006]{2006Scharmer}
Scharmer, G. (2006).
\newblock {Comments on the optimization of high resolution Fabry-P{\'{e}}rot
  filtergraphs}.
\newblock {\em Astronomy {\&} Astrophysics}, 447:1111--1120.

\bibitem[Scharmer et~al., 2003]{Scharmer2003}
Scharmer, G.~B., Bjelksj{\"{o}}, K., Korhonen, T., Lindberg, B., and Petterson,
  B. (2003).
\newblock {The 1-meter Swedish solar telescope}.
\newblock {\em Proc. SPIE}, 4853(August):341--350.

\bibitem[Scharmer et~al., 2008]{Scharmer2008}
Scharmer, G.~B., Narayan, G., Hillberg, T., {de la Cruz Rodriguez}, J.,
  L{\"{o}}fdahl, M.~G., Kiselman, D., S{\"{u}}tterlin, P., van Noort, M., and
  Lagg, A. (2008).
\newblock {CRISP Spectropolarimetric Imaging of Penumbral Fine Structure}.
\newblock {\em The Astrophysical Journal}, 689(1):L69--L72.

\bibitem[Schmidhuber, 2015]{Schmidhuber2015}
Schmidhuber, J. (2015).
\newblock {Deep Learning in Neural Networks: An Overview}.
\newblock {\em Neural Networks}, 61:85--117.

\bibitem[Sim{\~{o}}es et~al., 2017]{Simoes2017}
Sim{\~{o}}es, P. J.~A., Kerr, G.~S., Fletcher, L., Hudson, H.~S.,
  {Gim{\'{e}}nez de Castro}, C.~G., and Penn, M. (2017).
\newblock {Formation of the thermal infrared continuum in solar flares}.
\newblock {\em Astronomy {\&} Astrophysics}, 605:A125.

\bibitem[Skumanich and Lites, 1987]{Skumanich1987}
Skumanich, A. and Lites, B.~W. (1987).
\newblock {Stokes Profile Analysis and Vector Magnetic Fields I. Inversion of
  Photospheric Lines}.
\newblock {\em The Astrophysical Journal}, 322:473--482.

\bibitem[{Socas-Navarro} et~al., 2015]{Socas-Navarro2014}
{Socas-Navarro}, H., {de la Cruz Rodr{\'{\i}}guez}, J., {Asensio Ramos}, A.,
  {Trujillo Bueno}, J., and {Ruiz Cobo}, B. (2015).
\newblock {An open-source, massively parallel code for non-LTE synthesis and
  inversion of spectral lines and Zeeman-induced Stokes profiles}.
\newblock {\em \aap}, 577:A7.

\bibitem[Socas‐Navarro et~al., 2000]{SocasNavarro2000}
Socas‐Navarro, H., {Trujillo Bueno}, J., and {Ruiz Cobo}, B. (2000).
\newblock {Non‐LTE Inversion of Stokes Profiles Induced by the Zeeman
  Effect}.
\newblock {\em The Astrophysical Journal}, 530(2):977--993.

\bibitem[Sriperumbudur et~al., 2009]{Sriperumbudur2009}
Sriperumbudur, B.~K., Fukumizu, K., Gretton, A., Lanckriet, G. R.~G., and
  Sch{\"{o}}lkopf, B. (2009).
\newblock {Kernel choice and classifiability for RKHS embeddings of probability
  distributions}.
\newblock {\em Proceedings of the 23rd Annual Conference on Neural Information
  Processing Systems}, pages 1750----1758.

\bibitem[Tolstikhin et~al., 2017]{Tolstikhin2017}
Tolstikhin, I., Bousquet, O., Gelly, S., and Schoelkopf, B. (2017).
\newblock {Wasserstein Auto-Encoders}.
\newblock {\em ArXiv e-prints}, pages 1--18.

\bibitem[Uitenbroek, 2001]{Uitenbroek2001}
Uitenbroek, H. (2001).
\newblock {Multilevel Radiative Transfer with Partial Frequency
  Redistribution}.
\newblock {\em The Astrophysical Journal}, 557(1):389--398.

\bibitem[{Van Noort} et~al., 2005]{VanNoort2005}
{Van Noort}, M., {Van Der Voort}, L.~R., and L{\"{o}}fdahl, M.~G. (2005).
\newblock {Solar image restoration by use of multi-frame blind de-convolution
  with multiple objects and phase diversity}.
\newblock {\em Solar Physics}, 228(1-2):191--215.

\bibitem[Varady et~al., 2010]{Varady2010}
Varady, M., Ka{\v{s}}parov{\'{a}}, J., Moravec, Z., Heinzel, P., and
  Karlick{\'{y}}, M. (2010).
\newblock {Modeling of solar flare plasma and its radiation}.
\newblock {\em IEEE Transactions on Plasma Science}, 38(9 PART 1):2249--2253.

\bibitem[Vernazza et~al., 1981]{Vernazza1981}
Vernazza, J.~E., Avrett, E.~H., and Loeser, R. (1981).
\newblock {Structure of the Solar Chromosphere. III. Models of the EUV
  Brightness Components of the Quiet Sun}.
\newblock {\em The Astrophysical Journal Supplement Series}, 45(C):635--725.

\bibitem[Withbroe and Noyes, 1977]{Withbroe1977}
Withbroe, G. and Noyes, R. (1977).
\newblock {Mass and Energy Flow in the Solar Chromosphere and Corona}.
\newblock {\em Annual Review of Astronomy and Astrophysics}, 15:363--87.

\bibitem[Wulser and Marti, 1989]{Wulser1989}
Wulser, J. and Marti, H. (1989).
\newblock {High Time Resolution Observations of H$\alpha$ Line Profiles During
  the Impulsive Phase of a Solar Flare}.
\newblock {\em The Astrophysical Journal}, 341:1088--1096.

\end{thebibliography}
    \end{document}